\documentclass[sigconf, screen]{acmart} 
\usepackage[utf8]{inputenc} 
\usepackage[T1]{fontenc}    
\usepackage{amsmath}        
\usepackage{algorithm}      
\usepackage{algorithmicx}   
\usepackage{algpseudocode}
\usepackage{multirow}
\usepackage{array}
\usepackage{enumitem}
\setlist{noitemsep,parsep=0pt,partopsep=0pt, leftmargin=9pt} 
\usepackage{setspace} 
\usepackage{booktabs}
\usepackage{soul}
\AtBeginDocument{%
  }

\usepackage{xcolor}
\definecolor{color1}{RGB}{229, 148, 12}
\definecolor{color2}{RGB}{56, 152, 171}
\definecolor{color3}{RGB}{166,86,40}
\definecolor{color4}{RGB}{0, 85, 0}
\definecolor{color5}{RGB}{0, 0, 0}

\newcommand{\eg}{{\it e.g.,\ }}

\newcommand{\ie}{{\it i.e.,\ }}

\newcommand{\q}[1]{\textit{``#1''}}

\newcommand{\rws}[1]{\textcolor{color1}{#1}}

\newcommand{\rv}[1]{\textcolor{color5}{#1}}

\newcommand{\triple}{triple}
\newcommand{\triples}{triples}

\newcommand{\cpanel}{LLM Conversation Panel}

\newcommand{\tool}{NeuroSync~}
\newcommand{\toole}{NeuroSync}
\setcopyright{acmlicensed}
\copyrightyear{2025}
\acmYear{2025}
\acmDOI{10.1145/3746059.3747668}
\acmConference[UIST '25]{The 38th Annual ACM
Symposium on User Interface Software and Technology}{September 28-October
1, 2025}{Busan, Republic of Korea}
\acmBooktitle{The 38th Annual ACM Symposium on User Interface Software
and Technology (UIST '25), September 28-October 1, 2025, Busan, Republic of Korea}
\acmISBN{979-8-4007-2037-6/2025/09}

\begin{document}


\title[\toole]{\toole: Intent-Aware Code-Based Problem Solving via\\ Direct LLM Understanding Modification}



\author{Wenshuo Zhang}
\affiliation{%
  \institution{The Hong Kong University of Science and Technology}
  \city{Hong Kong SAR}
  \country{China}}
\orcid{0009-0007-9226-0713}
\email{wzhangeb@connect.ust.hk}

\author{Leixian Shen}
\affiliation{%
  \institution{The Hong Kong University of Science and Technology}
  \city{Hong Kong SAR}
  \country{China}}
\orcid{0000-0003-1084-4912}
\email{lshenaj@connect.ust.hk}

\author{Shuchang Xu}
\affiliation{
\institution{The Hong Kong University of Science and Technology}
\city{Hong Kong SAR}
\country{China}}
\orcid{0000-0002-7642-9044}
\email{sxuby@connect.ust.hk}

\author{Jindu Wang}
\affiliation{
\institution{The Hong Kong University of Science and Technology}
\city{Hong Kong SAR}
\country{China}}
\orcid{0009-0009-4028-4662}
\email{jwangki@connect.ust.hk}

\author{Jian Zhao}
\affiliation{
\institution{University of Waterloo}
\city{Waterloo}
\country{Canada}}
\orcid{0000-0001-5008-4319}
\email{jianzhao@uwaterloo.ca}

\author{Huamin Qu}
\affiliation{
\institution{The Hong Kong University of Science and Technology}
\city{Hong Kong SAR}
\country{China}}
\orcid{0000-0002-3344-9694}
\email{huamin@ust.hk}

\author{Lin-Ping Yuan}
\affiliation{
\institution{The Hong Kong University of Science and Technology}
\city{Hong Kong SAR}
\country{China}}
\authornote{Corresponding author.}
\orcid{0000-0001-6268-1583}
\email{yuanlp@cse.ust.hk}

\renewcommand{\shortauthors}{Zhang et al.}

\begin{abstract}

Conversational LLMs have been widely adopted by domain users with limited programming experience to solve domain problems. 
However, these users often face misalignment between their intent and generated code, resulting in frustration and rounds of clarification.
This work first investigates the cause of this misalignment, which dues to \textit{bidirectional ambiguity}: both user intents and coding tasks are inherently nonlinear, yet must be expressed and interpreted through linear prompts and code sequences. 
To address this, we propose \textit{direct intent–task matching}, a new human–LLM interaction paradigm that externalizes and enables direct manipulation of the \textit{LLM understanding}, i.e., the coding tasks and their relationships inferred by the LLM prior to code generation.
As a proof-of-concept, this paradigm is then implemented in \textit{NeuroSync}, which employs a knowledge distillation pipeline to extract LLM understanding, user intents, and their mappings, and enhances the alignment by allowing users to intuitively inspect and edit them via visualizations. 
We evaluate the algorithmic components of NeuroSync via technical experiments, and assess its overall usability and effectiveness via a user study \rv{(N=12)}.
The results show that it enhances intent–task alignment, lowers cognitive effort, and improves coding efficiency.

\end{abstract}






\maketitle

\section{Introduction}

Programming is an essential and practical tool for domain users to solve problems within their areas of expertise. 
For instance, a marine biologist might need to analyze large amounts of ocean data to study climate change or marine ecosystems.
These domain users often lack programming skills and struggle to implement these solutions themselves.
Conversational Large Language Models (LLMs), such as ChatGPT, have become popular among these users~\cite{waitgpt} because they allow users to express their problem-solving intents through natural language prompts and receive automatically generated code. This lowers the barriers to leveraging programming to solve problems.
Despite this benefit, users frequently encounter \textit{misalignment} between their intents and the code generated by LLMs~\cite{Gulf_of_Envisioning}. This misalignment typically leads to repetitive cycles of clarification and debugging, causing frustration and task failure.

Current efforts addressing misalignment generally fall into two categories.
The first body of work focuses on improving \textit{user-to-LLM communication}, aiming to help users formulate clear and structured prompts, such as logically organized coding tasks or pseudo-code~\cite{CoLadder,aichain}.
The second targets \textit{LLM-to-user communication}, which seeks to enhance users' understanding of generated code via interactive explanations and visualizations~\cite{waitgpt,Ivie}. 
While effective, they are primarily designed to support professional programmers, who have the expertise to decompose problems into coding tasks and interpret the generated code.
In contrast, domain users lack the expertise to identify or articulate misalignment through direct interaction with code. They instead rely heavily on conversational interactions with LLMs, and often result in notable friction to complete their tasks with current approaches.

\rv{To better understand why misalignment exists during the conversational process and how it can be effectively addressed}, we conducted a two-phase formative study with six domain users to investigate this problem.
In the first phase, we analyzed human-LLM conversation histories arising from their daily work.
We uncovered \textit{bidirectional ambiguity} as a key reason for misalignment: both user intents and coding tasks are inherently nonlinear and dynamic, yet must be communicated through linear prompts and code representations.
Building on this insight, the second phase centered on finding suitable visual representations for alleviating such misalignment. We explored graph visualizations for non-linear coding tasks and employed a tech probe to evaluate their pros and cons.

\begin{figure}[t]
    \centering
    \includegraphics[width=\linewidth]{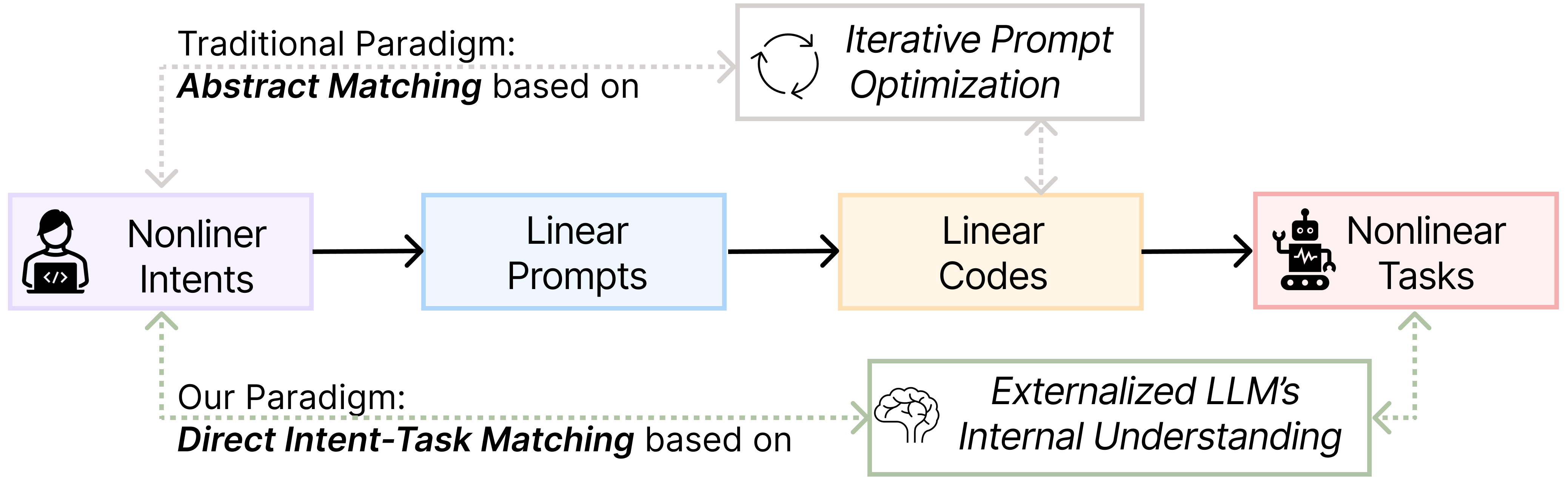}
    \caption{Comparison between the proposed \textit{direct intent–task matching} paradigm and the traditional paradigm for user-LLM interactions in programming.}
    \label{fig:concept_overview}
\end{figure}

Upon deeper understanding of the \textit{bidirectional ambiguity}, we propose a novel human-LLM interaction paradigm called \textit{direct intent–task matching} to address this issue (Fig.~\ref{fig:concept_overview}).
In traditional paradigms, LLMs generate code directly from user prompts without revealing their ``internal understanding''. 
In contrast, our approach introduces a transparent process that externalizes the \textit{LLM understanding}, which refers to the coding tasks and their relationships. These tasks and relationships are inferred by the LLM from a user’s prompt and serve as the basis for the code it generates.
This paradigm allows users to directly interact with the \textit{LLM understanding}, diagnosing and correcting any inaccuracies or misalignment, before code is generated. 

\begin{figure}
    \centering
    \includegraphics[width=\linewidth]{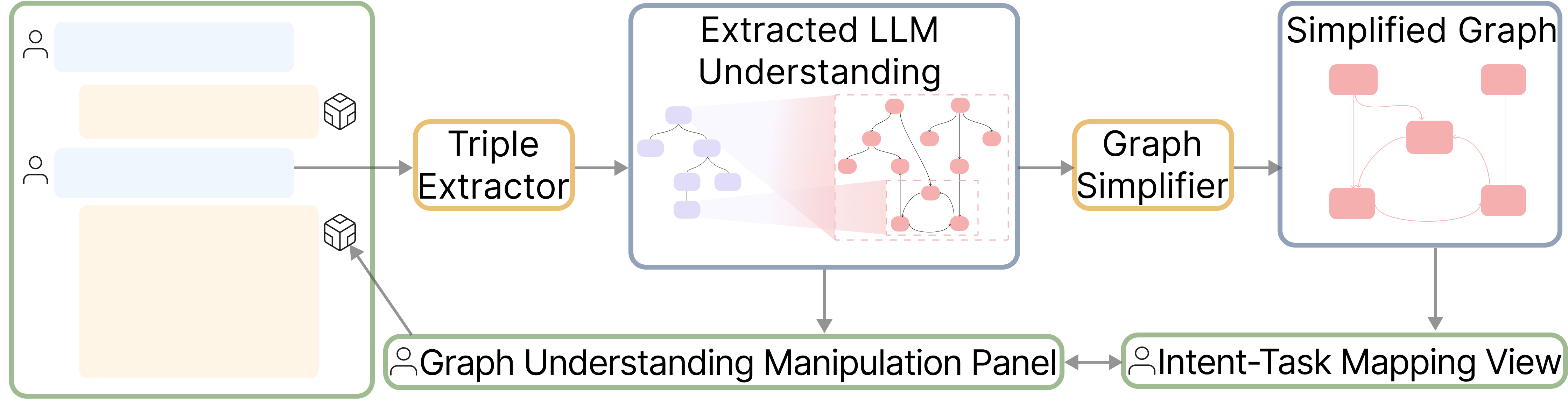}
    \caption{Overview of \toole, a proof-of-concept implementation of the \textit{direct intent–task matching} paradigm. \tool takes user prompts as input, extracts the \textit{LLM understanding}, enables users to refine this understanding through graph-based visualizations, and feeds the refined understanding back to the LLM to generate code that more accurately aligns with user intents.}
    \label{fig:system_overview}
\end{figure}

We operationalized this paradigm in a proof-of-concept system named \toole.
As shown in Fig.~\ref{fig:system_overview}, when a user inputs a prompt, \tool first extracts the LLM understanding, user intents, and their mappings.
It then visualizes the LLM understanding as a graph and user intents as a tree, 
allowing users to manipulate and refine the graph according to their intents visualized in the tree. Once the user confirms, \tool generates and displays code guided by the updated LLM understanding, ensuring it accurately aligns with the user intents.
Furthermore, \tool incorporates two algorithmic components to address two critical challenges during the process.
First, users often lose track of how the LLM understanding graph evolves while the intent changes, especially when the graph grows more complex. 
We thus designed an intent-aware graph simplication algorithm that can identify the nodes related to the intent changes, allowing for emphasizing them in the visualization.  
Second, extracting LLM understandings and user intents can be computationally heavy. We then leveraged a novel distillation pipeline to fine-tune small language models to enable faster extraction.

We carried out technical experiments to assess the graph implications algorithm and the distillation pipeline, revealing the effectiveness of our designed algorithms. 
We further evaluated \tool through a controlled user study with 12 domain users, compared to a customized baseline. 
The results demonstrate that \tool \rv{substantially} improves intent–task alignment, reduces cognitive load, and enhances coding efficiency, confirming the effectiveness of \textit{direct intent–task matching} as a novel conversational coding paradigm for non-professional programmers.

In summary, our contributions are threefold:
\begin{itemize}
    \item We introduce \textit{direct intent–task matching}, a paradigm that externalizes \textit{LLM understanding}, enabling domain users with limited programming expertise to refine and align coding tasks with their intents before code generation.
    \item We develop \toole, a proof-of-concept system that implements this paradigm with interactive graph-based visualizations and two algorithmic components: a simplification algorithm to manage graph complexity and a distillation pipeline for faster extraction of LLM understanding.
    \item We validate \tool through technical experiments and a user study, demonstrating its ability to improve intent–task alignment, reduce cognitive load, and enhance coding efficiency.
\end{itemize}

\begin{table*}[h]
\centering
\small
\caption{Demographics of Participants in the Formative Study. `Exp.' denotes their user experience within their respective domains. `Task' indicates the specific tasks performed according to their provided dialog history. `Rounds' represents the number of conversation rounds conducted.}
\label{tab:user_info}
\setlength{\tabcolsep}{2pt}
\begin{tabular}{p{0.3cm}p{3cm}p{1.2cm}p{6.3cm}p{4.5cm}p{1.cm}}
\hline
\textbf{ID} & \textbf{Domain} & \textbf{Exp.} & \textbf{Task} & \textbf{Interaction Mode/LLM Name} & \textbf{Rounds} \\
\hline
P1 & Design (Ph.D.) & 1.5 Years &  Web article crawlers with python & Coding Mode/Qwen & 11 \\

P2 & Electric Grid (UG) & 0.5 Years & Signal analysis with Matlab & Chatting Mode/Doubao  & 13 \\

P3 & Clinical Medicine (Ph.D.) & 6 Years & Medical image processing and classification with Python & Chatting Mode/Kimi & 21 \\

P4 & Theory Math (Ph.D.) & 3 Years &  Table drawing and adjusting with Latex & Coding Mode/Qwen & 8 \\

P5 & Policy Making (M.S.) & 1 Year & Interview data analysis and visualization with python & Coding Mode/GPT-4o & 13 \\

P6 & Economics (M.S.) & 5 Years & Financial data analysis and visualization with python & Data Analysis Mode/GPT-4o & 29 \\
\hline
\end{tabular}
\end{table*}

\section{Related Work}

\subsection{Human–LLM Alignment in Coding}

Human–LLM alignment seeks to ensure that LLM outputs reflect users’ intentions, particularly in code-based problem-solving scenarios~\cite{align_survey_all, align_survey_ai}. Two key issues are commonly identified: how to help users formulate effective instructions, and how to assess whether the generated outputs match their intents~\cite{align_survey_all_HCI}.

Prompt engineering is the primary approach for improving alignment in conversational coding systems. Sarkar et al.~\cite{sarkar2022like} highlight users’ iterative refinement of prompts, describing this iterative process as \textit{abstract matching}, which reflects Norman's gulf of execution~\cite{glufofexecution}. In addition to this, users also encounter a related cognitive barrier known as the ``gulf of envisioning''~\cite{Gulf_of_Envisioning}, where articulating clear tasks and anticipating model outputs is challenging. To bridge this cognitive gap, prior works have proposed various strategies such as in-context learning~\cite{icl_survey, icl_code}, structured decomposition approaches (\eg AI Chains~\cite{aichain}, CoLadder~\cite{CoLadder}), and the \textit{feedforward} mechanism~\cite{min2025feedforward, feedforw13, Instructions}. \rv{Among them, CoLadder~\cite{CoLadder} uses unidirectional, monologue-driven interaction and content in ``block'' or ``chain'' UIs~\cite{aichain} are generated statically rather than automatically adapting to users' changing intent during usage, while NeuroSync is used in a bidirectional dialogue scenario and generates editable, simplified graphs on the fly based on dynamic user intents in each interaction round.}

\rv{Feedforward specifically anticipates outcomes before actions occur by prompting the model to generate interactive previews, such as predicted visual outputs, code suggestions, or descriptive summaries, based on user input. These previews help minimize unnecessary clarification rounds, bridge abstraction gaps between users and AI~\cite{abstraction_gap, SQLucid, gulf_evaluation}, and support users’ metacognitive reasoning~\cite{tankelevitch2024metacognitive}.}
Compared to existing work~\cite{abstraction_gap} that directly presents real LLM-generated code directly as feedforward information~\cite{abstraction_gap}, \tool externalizes the LLM's internal understanding---the tasks and their relationships that are implicitly encoded in the generated code---before code is generated. This reduces the cost of iteration and ensures consistency between preview and final output via knowledge distillation~\cite{KD}.

\rv{Alignment and intent specification challenges are also well-studied in program synthesis, where the focus has shifted from complete formal specifications to interpreting ambiguous, high-level user intents. To address this ambiguity, interactive methods refine intent through user feedback, such as selecting distinguishing inputs to disambiguate candidate programs~\cite{zhang2020interactive, le2017interactive}, or through multi-turn conversational dialogues~\cite{nijkamp2022codegen}. For complex tasks, Gulwani’s work in programming-by-example introduced divide-and-conquer strategies~\cite{gulwani2011automating}, recursively partitioning examples, synthesizing distinct programs for subsets, and composing results with conditionals. More recently, reinforcement learning has been used to align programs with functional correctness by rewarding models for passing unit tests~\cite{Yu2023BCoderVD}.
A key distinction lies in the abstraction level: traditional pre-LLM methods emphasize code-level interactions (\textit{e.g.,} input-output examples~\cite{gulwani2011inputoutputexample, pbe2}, execution traces~\cite{trace, trace2}, or program sketches~\cite{solar2006sketch}), whereas \tool operates at the task level, externalizing high-level LLM tasks before code generation.}

Current research also highlights the cognitive load associated with comprehending and debugging LLM-generated code~\cite{load1,load2,CoLadder}. Approaches such as real-time explanations~\cite{Live_Programming,Ivie} and visualization-based presentations~\cite{waitgpt} have been proposed to ease understanding. Unlike these post-generation methods, \tool intervenes before code generation, explicitly visualizing tasks and their relations, thus reducing cognitive complexity at an earlier stage.

\subsection{Graph-Based Interfaces for LLM Interaction}

Conversational LLM interfaces predominantly adopt linear interaction flows, where user prompts and model responses are serialized in turn~\cite{waitgpt}. While simple, such interfaces struggle with complex tasks due to high cognitive load~\cite{waitgpt}, versioning issues~\cite{cells_for_writting}, and limited user control~\cite{memolet, directgpt}. 
To overcome these limitations, recent systems explore alternative interaction paradigms for complex tasks like creative design~\cite{luminate}, exploratory analyses~\cite{sensecape}, and data analysis programming~\cite{waitgpt}.

Graphs naturally represent relational structures via nodes and edges~\cite{graph}, with diverse visualizations like node-link diagrams~\cite{diagrams, GEGraph} and hierarchical trees~\cite{mindaloguetree}. Recent studies have leveraged graph-based UIs to enhance LLM-driven management and logical reasoning~\cite{yan2024knownet}. For instance, Kim et al.~\cite{cells_for_writting} abstract writing processes into graphs to facilitate versioning; WaitGPT~\cite{waitgpt} visualizes analysis tasks via data-flow graphs; Promptchainer~\cite{promptchainer} and GoT~\cite{GoT} use graph representations to support multi-step task decomposition. Moreover, tools like Low-code LLM~\cite{Low_code_llm} and CoLadder~\cite{CoLadder} employ graph interfaces to directly organize intents and code snippets, improving alignment and clarity.

Recognizing these advantages, \tool adopts graph-based visualizations to represent coding tasks and their relations. While existing studies focus mostly on static graph visualizations and cognitive load management~\cite{Graph_Cogn}, \tool introduces dynamic graph simplification that automatically adapts graphs in real-time according to evolving user intents, reducing users' cognitive burden during multi-turn interactions.

\subsection{LLM Reasoning and Task Structuring}

Recent advances in applying LLMs to reasoning-intensive tasks (\eg programming~\cite{cot_code}, mathematics~\cite{mathLLM}) have identified limitations in reasoning capabilities of basic LLM models~\cite{not_reseaning}. To enhance reasoning, researchers proposed techniques like prompt tuning~\cite{prompt_learning, prompt_survey} and structured reasoning prompts such as Chain-of-Thought (CoT)~\cite{CoT}, Least-to-Most (L2M)~\cite{LtM}, and Tree-of-Thought (ToT)~\cite{tot}. These methods decompose reasoning into intermediate steps, enhancing model performance on complex problems.

CoT decomposes complex problems into linear sub-steps, which is suitable for problems with sequential logic but has limited ability to support nonlinear reasoning~\cite{bi2024forest}. In contrast, methods such as L2M~\cite{LtM} incrementally introduce complexity, avoiding premature limitations in thought exploration. Variants like Path-of-Thought (PoT)~\cite{PoT}, Concept Composition (CoC)~\cite{CoC}, and Aggregation-of-Thought (AoT)~\cite{AoT} further optimize efficiency and performance. Meanwhile, ToT explores choices via tree-structured decision points, supporting complex multi-step tasks, with Graph-of-Thoughts (GoT)~\cite{GoT} providing an aggregated state exploration approach.

These approaches aim to improve internal LLM reasoning. In contrast, \tool externalizes and exposes the LLM’s inferred task structure---what we term the model’s \textit{internal understanding}---before execution. While different in purpose, both lines of work reflect the importance of intermediate structure in aligning model behavior with user intent. In \toole, this structure serves not only as guidance for code generation but also as a manipulable medium for user–LLM alignment.


\section{Formative Study}
\label{Sec: Formative}

To examine the causes of human–LLM misalignment in conversational coding and inform system design, we conducted a formative study with domain users who have little coding experiences. The study included (1) interaction history analysis and interviews to uncover misalignment patterns, and (2) semi-structured interviews, informed by a literature review, to explore effective representations of graphs for conveying code tasks and user intent.

\subsection{Study 1: Understanding Human-LLM Misalignment}

\subsubsection{Participants and Data.} We invited six participants (P1- P6) to conduct retrospective analysis. They were from diverse domain backgrounds with varying levels of programming experience and education background (Tab.~\ref{tab:user_info}). 
Each provided full records of prior real-world interactions with LLMs while performing coding tasks for problem-solving. On average, each task had 15.83 interaction rounds (SD = 7.76) per session, spanning four LLM platforms and multiple programming languages (\eg Python, MATLAB, LaTeX) under different LLM modes.

To analyze the reasons behind misalignment during conversational code generation, we conducted an analysis on users' interaction history and observed that the phenomenon of LLMs failing to accurately generate code aligned with user intent is widespread (6/6) and leads to a number of useless interactions. 



\subsubsection{Analysis Protocol.}
\rv{We analyzed participants' interaction histories with LLMs using an open-coding approach~\cite{opencode}. Two authors independently coded the data and iteratively refined the codes until reaching agreement. The codes were then grouped into themes, which were carefully reviewed and discussed to identify the key findings of the study.}

We first annotated each interaction round with:
(1) \textit{User intent}, inferred through participant clarification and interaction review;
(2) \textit{Prompt quality}, evaluated by independent raters to assess how well the prompt conveyed the intended task;
(3) \textit{LLM-executed tasks}, extracted from generated code and mapped to user intent.

Building on these annotations, we further identified potential misalignment points—rounds where user intent remained stable but the generated code deviated from the intended task. This process allowed us to trace breakdowns across turns and identify underlying causes of intent–task misalignment.

\begin{figure}
    \centering
    \includegraphics[width=\linewidth]{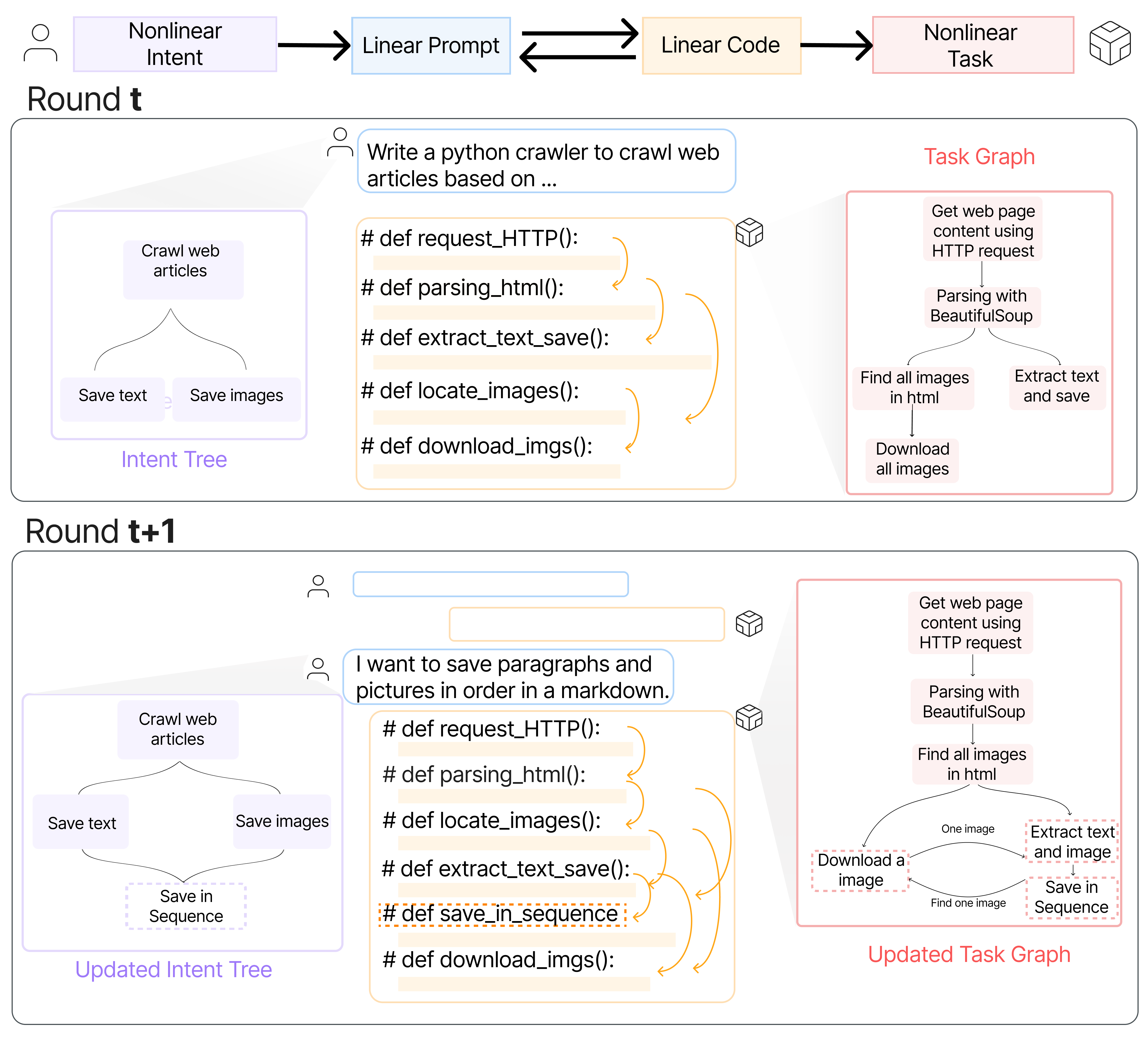}
    \caption{Illustration of bidirectional ambiguity.}
    \label{fig:formative_example}
\end{figure}

\subsubsection{Findings}

\textbf{Bidirectional ambiguity is a major cause of human-LLM misalignment in conversational coding tasks.}  
During the conversation with LLM for coding tasks, ambiguity is bidirectional:  
\textit{User-to-LLM}: Users find it challenging to clearly express their needs and the information required by the LLM in their prompts. For example, converting tree-like intent in Fig.~\ref{fig:formative_example} into prompt will lose direct structure and cause ambiguity.
\textit{LLM-to-User}: Users struggle to understand the specific tasks and execution logic embedded in the code, making it difficult to provide precise modification requests. For example, in Fig.~\ref{fig:formative_example}, users need to reconstruct codes and code relationships by themselves, which is difficult and low ability of understanding code will lead to ambiguity.
This bidirectional ambiguity compounds over turns, causing LLMs to produce code misaligned with user intent. As LLM capabilities grow and inference slows, the cost of these ineffective interactions increases.


\textbf{User-to-LLM Ambiguity.}
Users often struggle to express their intent clearly due to three key issues:
(1) \textit{Nonlinear intent loss}: User goals are typically hierarchical and evolve over time. However, when mapped into linear prompts, this structure is flattened, leading to semantic ambiguity and loss of global intent.
(2) \textit{Contextual omissions}: Prompts often lack critical information due to delayed articulation and limited user memory. LLMs, with constrained context windows, may miss important prior requirements.
(3) \textit{Vague modification guidance}: Domain users, unfamiliar with code internals, frequently describe desired outcomes without specifying what to change. This hampers LLMs to revise code accurately.

\textbf{LLM-to-User Ambiguity.}
LLM-generated code embeds multiple interrelated tasks, often structured nonlinearly. Users with limited programming experience face difficulty in unpacking this structure, identifying task boundaries, and understanding the model’s reasoning. As a result, they may overlook unintended logic, misinterpret execution flow, or fail to spot partial completions, making it hard to issue precise follow-up instructions. This impairs both comprehension and correction, especially in multi-round interactions where misunderstandings accumulate.

\subsubsection{Implications} 
To reduce LLM-to-user ambiguity, systems must present model-inferred tasks in a more interpretable form, allowing users, especially those with limited coding expertise, to understand and guide code generation without reading raw code.

\begin{table}[t]
\centering
\small
\caption{Comparison of different task representations.}
\label{tab:representation_methods}
\renewcommand\arraystretch{1}
\setlength{\tabcolsep}{1mm}{
\begin{tabular}{m{3.8cm}m{1.1cm}m{1cm}m{1cm}}
\hline
\textbf{Representation Method} & \textbf{Intuitive} & \textbf{Easy to Modify} & \textbf{Layman-friendly} \\
\hline
Task Flow Diagram (Graph)\rv{~\cite{graph_survey}}  & $\checkmark$ & $\checkmark$ & $\checkmark$ \\

Pseudocode\rv{~\cite{sudocode}} & $\times$ & $\times$ & $\times$ \\

UML Activity Diagram\rv{~\cite{dumas2001uml}} & $\checkmark$ & $\checkmark$ & $\times$ \\

Decision Table\rv{~\cite{pooch1974translation}}  & $\times$ & $\times$ & $\times$ \\

Data Flow Diagram\rv{~\cite{waitgpt}}  & $\checkmark$ & $\checkmark$ & $\times$ \\

Natural Language Description & $\times$ & $\times$ & $\checkmark$ \\
\bottomrule
\end{tabular}}

\end{table}

\subsection{Study 2: Exploring Graph-Based Representations for Code Tasks}
As discovered in Study 1, prompts and code presented in linear forms do not effectively convey non-linear intents or help users understand non-linear code tasks involved in multiple rounds of interactions. 
Therefore, in Study 2, we aim to explore whether there are better representations that improve the communication of non-linear coding tasks between users and LLMs.

\subsubsection{Representation Survey.}  
\rv{We first conducted an expert-driven ideation to enumerate common formats, and then verified each through a focused review of recent representative papers. This approach allowed us to capture a broad spectrum of practical representations without relying on unfocused keyword searches. Results are shown in Tab.~\ref{tab:representation_methods}.
} Among these, graph-based structures (\eg task flow graphs) emerged as intuitive, editable, and directly tied to task logic. While promising, we also observed that as programming intent evolves, task graphs can become increasingly complex, highlighting the need for intent-aware abstraction and scalable visual structures.


\begin{figure}
    \centering
    \includegraphics[width=\linewidth]{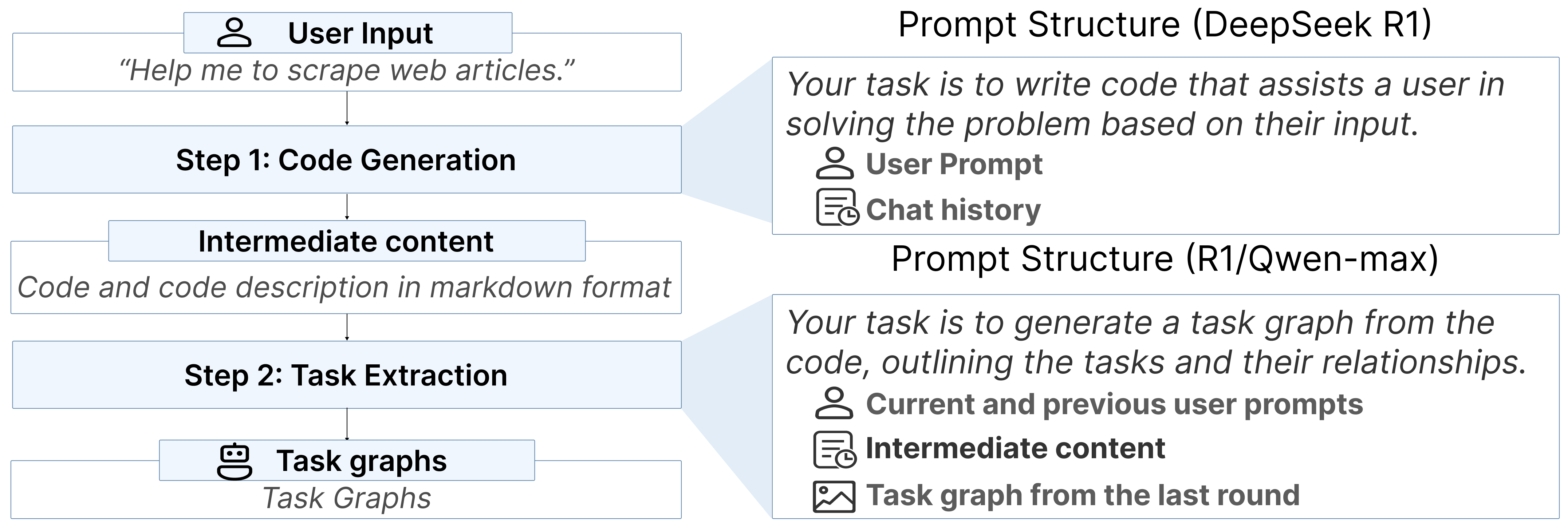}
    \caption{Two-stage extractor. Graph-based task representation will be extracted after the code is generated based on the user prompt.}
    \label{fig:probe}
\end{figure}

\subsubsection{Graph-Based Two-stage Extractor.}  
To evaluate the feasibility of graph-based representations in real-world interactions, we developed a prototype tool that extracts task graphs from user–LLM interactions. The probe processes both initial prompts and follow-up inputs to incrementally build and update task-level representations (Fig.~\ref{fig:probe}). It supports:

\begin{itemize}
    \item \textit{Task Graph Initialization:} At the first interaction, the probe analyzes the user's prompt and the LLM's response (codes) to identify and structure basic tasks, subtask groups, and task dependencies behind the codes. The result is a hierarchical graph, where nodes represent tasks and edges denote logical or sequential relationships. 
    
    \item \textit{Task Graph Refinement:} In subsequent turns, the graph is incrementally updated based on new prompts, responses, and historical context. Refinement includes task additions, deletions, and updates, especially around variable usage and logic changes, to ensure consistency and traceability across rounds.
\end{itemize}

Task graphs are rendered as part of the conversation interface and stored for further review. However, during early trials, we noted that API latency became a bottleneck in multi-turn settings, pointing to the need for lighter-weight implementations.

\subsubsection{Study Protocol.}  
To assess the effectiveness of task graphs, we conducted semi-structured interviews with the same six interdisciplinary users in Study 1 (P1-P6)\rv{~and used an open-coding approach~\cite{opencode} for data analysis. }
We first extracted samples from three key stages of user–LLM interaction: initialization, intent progression, and completion. Each sample included the prompt, LLM-generated code, and the corresponding task graph, presented sequentially to participants. The procedure includes three stages.
In Stage 1, participants reviewed only the generated code and described their understanding and obstacles.
In Stage 2, they examined both the code and task graph, comparing the ease of task comprehension with and without the graph, and offered suggestions for improvement.
In Stage 3, we showed consecutive code–graph pairs from two interaction turns and asked participants to identify task changes, misunderstandings, and strategies for tracking differences.
Interviews focused on two core questions:  
(1) Do graphs improve task understanding compared to directly reading code?  
(2) What features are needed to support effective graph interpretation and updates?
Throughout the process, we observed participants’ behaviors in navigating graph changes and concluded with a reflection session to gather feedback on cognitive load, update clarity, and expectations for graph design.


\subsubsection{Findings}  
All participants reported difficulty in understanding raw code due to limited programming knowledge. Key barriers included misalignment between linear code flow and branching task logic (P1, P4, P5), and high cognitive load from memorizing variable usage and logic transitions (P3, P6). In contrast, task graphs were consistently viewed as more helpful. Participants highlighted two key benefits:  
(1) Improved task comprehension through clear visualization of task dependencies and subgoals (P1, P3, P4, P5, P6);  
(2) Enhanced efficiency in locating key logic points and understanding overall code purpose (P1, P3, P5).

However, as interaction rounds increased, graph complexity grew and negatively impacted interpret ability. While some participants were able to reconstruct evolving task intents (P1, P2, P5, P6), others reported confusion or errors (P3, P4). To address this, users suggested:
(1) Providing abstracted overviews with zoom-in capabilities (P1-P6);
(2) Supporting node-level exploration for localized inspection (P1, P4);
(3) Grouping and annotating task clusters to clarify hierarchy and dependencies (P2, P3, P5, P6).



\subsubsection{Implications.}  
Use graph-based representations to externalize LLM coding tasks and tree structures to model user intents, enabling fine-grained alignment. To ensure responsiveness, employ lightweight feedforward representations and dynamically abstract graph complexity based on evolving user intent.

\subsection{Design Considerations}
Based on the findings from the two phases, we derive several design considerations:

\textbf{DC1: Support Bidirectional Disambiguation through Intent and Task Externalization.}  
To mitigate user–LLM misalignment, systems should support bidirectional disambiguation. On the user-to-LLM side, this requires making user intent explicit and editable, preserving task structure and global context beyond linear prompt input. On the LLM-to-user side, the model’s inferred coding tasks should be externalized in interpretable forms, enabling non-programmers to understand, verify, and adjust the system’s interpretation without reading raw code.

\rv{\textbf{DC2: Enhance Fluid Modification on Tasks to Align Intent in Multi-round Interactions.} Matching intent and tasks in conversational systems often requires multiple rounds, but high latency can disrupt users' focus, making it difficult to think fluidly and modify tasks effectively. To enable effective intent-task alignment,} systems should support low-latency interactions. However, extracting LLM understanding and user intent can be computationally expensive, especially when task structures grow in complexity. Therefore, the system should employ lightweight yet accurate feedforward mechanisms that allow for rapid generation and update of intermediate representations.




\textbf{DC3: Leverage Structured Graph Representations with Intent-Aware Abstraction.}  
Graph-based representations are effective for externalizing the LLM’s task structure, while user intent can be modeled as a tree, a specialized directed acyclic graph reflecting hierarchical goal decomposition. To manage complexity and cognitive load, systems should dynamically abstract or simplify task graphs based on evolving user intent, enabling scalable yet focused interaction across varying levels of detail.

\begin{figure*}[ht]
    \centering
    \includegraphics[width=\linewidth]{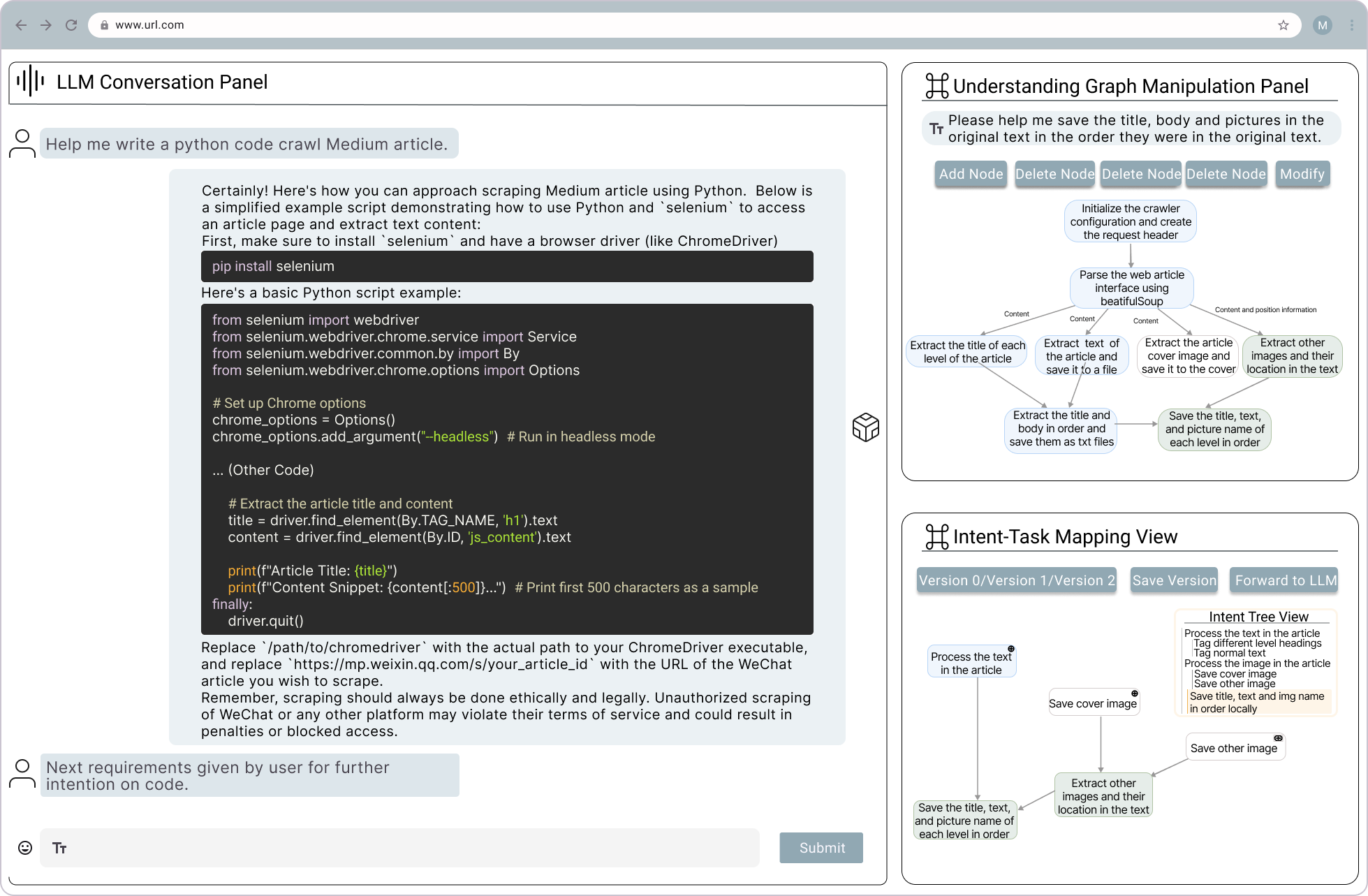}
    \caption{
Interface of \toole. Users interact with the LLM through Panel A (LLM Conversation Panel). Before each LLM response, the system generates an LLM understanding graph in Panel B (Understanding Graph Manipulation Panel) and a simplified version in Panel C (Intent–Task Mapping View). Users can edit the task graph in Panel B and explore task structures and intent alignment via Panel C.
}
    \label{fig:interface}
\end{figure*}

\section{Direct Intent-Task Matching}
Informed by \textbf{DC1}, we propose a novel human-LLM interaction paradigm called \textit{direct intent–task matching} (Fig.~\ref{fig:concept_overview}). The paradigm externalizes and enables direct manipulation of the \textit{LLM understanding} to support bidirectional disambiguation.


Specifically, \rv{inspired by the concept of \textit{user understanding}, where humans develop their interpretation of LLM outputs, we suggest that LLMs form a kind of understanding of user inputs. We call this \textbf{LLM understanding}}, which refers to the tasks and their relationships implicitly encoded in the code that an LLM is expected to generate based on user prompts.
By exposing this \textit{LLM Understanding} prior to code generation, users can interpret the intended tasks without reading raw code, reducing LLM-to-user ambiguity. 
Moreover, since the structure is editable, users can directly modify task representations, aligning them with their actual intent. This feedforward representation not only improves transparency but also serves as a lightweight, structured input alongside prompts, helping to resolve user-to-LLM ambiguity.

\textbf{Direct Intent–Task Matching} is a process that allows users to engage directly with the \textit{LLM understanding} before code is generated to address bidirectional ambiguity. Instead of relying on traditional prompt iteration or adjusting mismatched outputs after generation, users can iteratively refine how the LLM interprets their intent into specific coding tasks. This refinement resolves misalignments early in the process, ensuring the LLM's understanding evolves dynamically with each adjustment. By feeding this corrected understanding back into the LLM, users can achieve more efficient, interpretable, and intent-aligned code generation, streamlining the path from intent to output.

\begin{figure*}[ht]
    \centering
    \includegraphics[width=\linewidth]{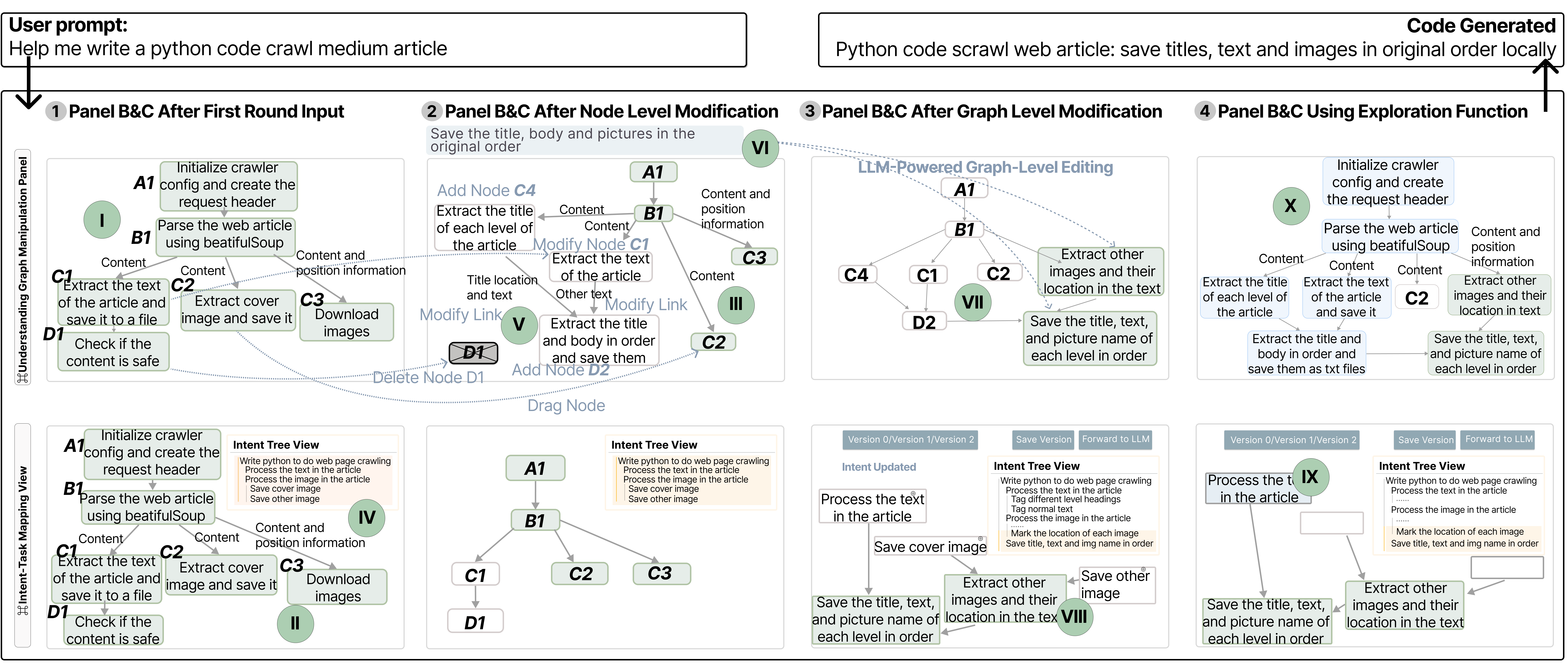}
    \caption{
User interactions with \toole. After entering a prompt in Panel A, users engage in a four-stage process of task exploration and modification in Panels B and C prior to code generation. 
(1) Upon prompt submission, an initial understanding graph is shown in Panel B, along with a simplified version in Panel C, where nodes associated with intent changes are highlighted. 
(2) Users can interactively explore the graph (\eg via dragging) and perform fine-grained node-level edits, such as modifying descriptions or adding nodes. 
(3) Alternatively, users may issue natural language commands to modify the graph. These updates are reflected in both panels, again with intent-relevant nodes highlighted. 
(4) Users may also click on merged nodes in Panel C to focus on corresponding subgraphs in Panel B. Once confirmed, the updated understanding is passed to the LLM for code generation. Selected nodes are zoomed in for clarity.    }
    \label{fig:usagecase}
\end{figure*}

\section{\tool}
We implement the \textit{direct intent–task matching} paradigm into a proof-of-concept system named \tool (Fig.~\ref{fig:system_overview}).

\subsection{Overview}
\tool operates in a multi-stage interaction loop. After the user submits a natural language prompt, \tool extracts a structured representation of the LLM understanding (\ie the predicted code-level tasks and their relationships), alongside the user’s intent and their mappings. These representations are then externalized in visual forms: the LLM understanding is shown as a graph based on our formative study\rws{~\ref{Sec: Formative}}, while the user intent is presented as a hierarchical tree structure based on previous research~\cite{CoLadder}. This visual design allows users to directly inspect, correct, and confirm task-level alignment, addressing \textbf{DC1} by supporting bidirectional disambiguation, as users no longer need to guess what the LLM will do, nor must they articulate intent solely through prompts.

To maintain low-latency interaction during multi-turn sessions, especially given the cost of extracting structured task representations, we implement a lightweight knowledge distillation pipeline\rws{~\ref{sec:triple_pipeline}}. This pipeline fine-tunes a small language model (SLM) to extract LLM understanding efficiently, based on training data generated by a multi-agent simulation system\rws{~\ref{sec:multiagent}}. This design meets \textbf{DC2}, ensuring that feedforward task information can be presented quickly and accurately without introducing delay or performance bottlenecks.

To support effective task visualization and user control, \tool adopts graph-based representations and integrates an \textit{intent-aware graph simplification algorithm}\rws{~\ref{sec:Simplification}} that dynamically adjusts the complexity of the understanding graph according to recent intent updates. This design directly responds to \textbf{DC3}, enabling scalable interaction through focused abstraction: users can access both global task structure and local task details, and selectively expand or highlight task components as needed.

\subsection{Usage Scenario}
\subsubsection{User Interface}
As shown in Fig.~\ref{fig:interface}, \toole’s user interface consists of (A) LLM Conversation Panel, (B) Understanding Graph Manipulation Panel, and (C) Intent-Task Mapping View.

\textbf{(A) LLM Conversation Panel.}  
It functions like a standard LLM interface, where users input prompts and receive responses. However, unlike traditional systems that rely solely on text prompts, \tool also incorporates the current version of the Understanding Graph into the generation process. Users can iteratively refine the graph without submitting a new prompt, enabling multiple rounds of graph-guided responses under the same user intent.

\textbf{(B) Understanding Graph Panel.}  
Before each LLM response, the system generates a task-level Understanding Graph based on the user’s prompt. Users can inspect and directly edit the graph to correct or clarify task understanding, including two levels of modification:

\begin{itemize}
    \item Graph-Level Modification: Users input natural language instructions into a \textit{modify block}, and \tool applies large-scale structural edits via the LLM API. This is ideal for reorganizing major task flows or introducing new task groups.
    \item Node-Level Modification: Users can manually adjust nodes and links, \ie adding, deleting, or editing task descriptions, enabling precise control over subtasks.
\end{itemize}


\textbf{(C) Intent–Task Mapping Panel.}  
To reduce cognitive load, this panel presents a simplified view of the Understanding Graph aligned with the user’s intent. It includes two components:

\begin{itemize}
    \item Intent Tree View: A hierarchical tree representing the user’s structured goals.
    \item Simplified Understanding Graph: A filtered version of the full graph, generated using our intent-aware graph simplification algorithm (Sec.~\ref{sec:Simplification}). It highlights nodes directly relevant to the current intent and merges irrelevant ones for clarity.
\end{itemize}

Panel C updates when a new prompt is submitted or when major edits occur in Panel B. Clicking a merged node highlights its corresponding region in the full Understanding Graph for focused inspection.

\subsubsection{Walkthrough Example}

Consider Kelly, a journalism student developing a web-based system for monitoring public opinion. To support her project, she needs a web crawler capable of extracting article content; however, she lacks the programming expertise to implement one herself. As a result, she turns to \tool for assistance.

Upon launching the LLM interface with \tool enabled, Kelly is presented with a standard conversation panel (see Fig.~\ref{fig:interface}A), accompanied by the Understanding Graph panel (B) and the Intent–Task Mapping panel (C). Kelly enters her request in the input field of Panel A, for example: \eg ``\textit{Help me write a Python script to crawl media articles}''. Unlike traditional LLM systems, \tool immediately visualizes its inferred understanding in Panel B (see I in Fig.~\ref{fig:usagecase}) and displays structured user intent in Panel C (II and IV in Fig.~\ref{fig:usagecase}). The system decomposes the request into a sequence of subtasks---such as ``\textit{Initialize the crawler configuration and create the request header}'' and ``\textit{Download images}''---and organizes them as a task flow diagram (see I in Fig.~\ref{fig:usagecase}). Panel C shows the same graph structure, highlighting all task nodes (see II in Fig.~\ref{fig:usagecase}) and presenting the overall intent tree (see IV in Fig.~\ref{fig:usagecase}).

Kelly interacts with the Understanding Graph by dragging nodes to explore its structure and quickly \textbf{identifies a misrepresentation of her original intent} (see III in Fig.~\ref{fig:usagecase}). Specifically, she notices that the graph includes a task related to content safety checking, which is outside the scope of her current objective. Instead, her goal is to distinguish between different title levels and save them alongside the main article text. To achieve this, she deletes the security-checking node and adds a new node labeled ``\textit{Extract the title at each level of the article}'' in Panel B. To accommodate this addition, she re-names the existing node ``\textit{Extract the text of the article and save it to a file}'' as ``\textit{Extract the title and body in order and save them as TXT files}''. She then creates appropriate links to integrate the new node into the existing task flow. These operations are shown in V in Fig.~\ref{fig:usagecase}.

After completing the initial modifications and finalizing the graph as shown in Panel B of Region 2, Kelly \textbf{encounters a new challenge}: she wants to save the titles, body text, and images from the original webpage in their original order. However, she finds this task complex and is uncertain how to proceed. She enters her revised requirement into the modification input block in Panel B (see VI in Fig.~\ref{fig:usagecase}) and clicks \textit{Modify}. In response, \tool automatically updates the Understanding Graph by adding two new nodes (highlighted in yellow in VII in Fig.~\ref{fig:usagecase}) and generates an updated, simplified graph along with a highlighted intent tree in Panel C (VIII in Fig.~\ref{fig:usagecase}). To reflect the refined goal, \tool merges relevant nodes in the Understanding Graph based on high-level intent tree components such as ``\textit{Process the text in the article}'' and ``\textit{Save cover image}'', selectively highlighting and retaining only the nodes directly associated with the updated intent.

Finding it challenging to interpret the entire Understanding Graph, Kelly clicks the extension button to focus on a specific sub-intent: ``\textit{Process the text in the article}'' (IX in Fig.~\ref{fig:usagecase}). In response, Panel B highlights the subset of nodes related to this intent, allowing her to concentrate more effectively on the text processing components (X in Fig.~\ref{fig:usagecase}). Once satisfied with the revised structure, Kelly clicks \textit{Confirm Graph}, prompting the LLM to generate code aligned with her refined understanding as represented in the graph.

Compared to her prior experiences---which typically required seven to eight rounds of back-and-forth interaction---Kelly is now able to complete the task in just one or two iterations. Minor errors are easily addressed through a quick update to the graph, followed by code regeneration.

\begin{figure}
    \centering
    \includegraphics[width=\linewidth]{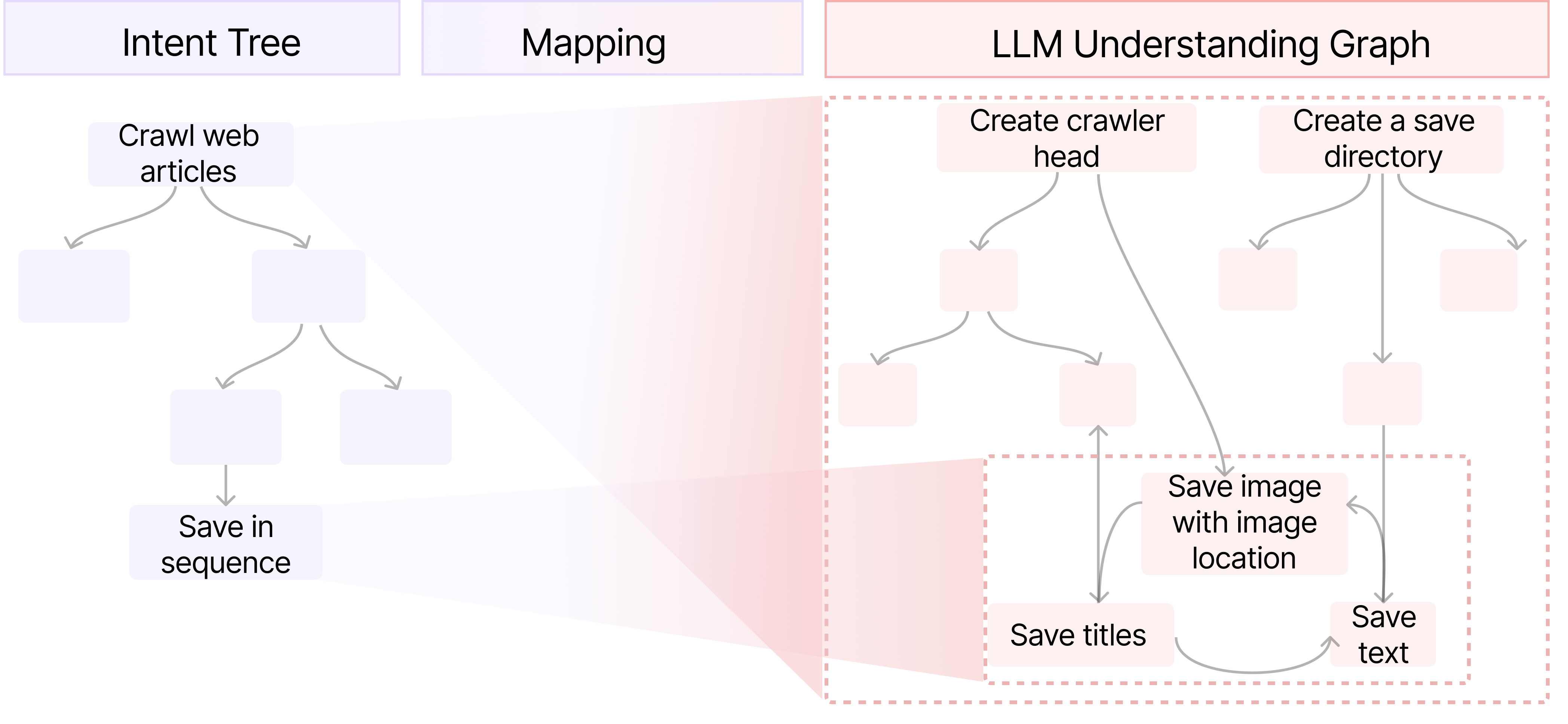}
    \caption{
Illustration of a \triple, which consists of an intent tree, an LLM understanding graph (\ie a task graph generated prior to code generation), and their mappings. Each node in the intent tree may correspond to one or more nodes in the LLM understanding graph. The figure highlights the mappings for two example intent nodes.
}
    \label{fig:triple}
\end{figure}

\subsection{Triple Extractor}

The Triple Extractor updates the LLM understanding, user intent, and their mappings each time a domain user inputs a prompt in the \cpanel. 
The extraction process must be efficient to reduce the delay perceived by users (\textbf{DC2}).

An intuitive but slow implementation is the two-stage extractor used in our formative study (Fig.~\ref{fig:probe}). 
While this approach achieves high accuracy, it requires two rounds of computationally expensive LLM calls and generates many intermediate tokens.
To improve efficiency, we construct the Triple Extractor using a fine-tuned Small Language Model (SLM) that can extract triples in a single step while maintaining sufficient similarity to those extracted by the two-stage extractor, \rv{\ie faithful reflection of intent, understanding and their mapping~\cite{faithful}}. This is achieved by a novel and cost-effective triple distillation pipeline (Fig.~\ref{fig:distillation}) that aligns the SLM with the LLM using data synthesized by a multi-agent system (Fig.~\ref{fig:playground}). 


\subsubsection{Triples}
For clarity, we first define a unified structure, referred to as triples, as follows:

\begin{equation}
    \label{eq:triple}
    \triples := \{ \texttt{Intent Tree}, \texttt{Understanding Graph}, \texttt{Mapping} \}
\end{equation}

\begin{itemize}
    \item \texttt{Intent Tree.} A hierarchical structure expressing the user's goal decomposition. The root node represents the high-level objective, while child nodes define sub-intents and operational details. This tree explicitly externalizes user intent in a form suitable for reasoning and alignment (Fig.~\ref{fig:triple} left).
    
    \item \texttt{Understanding Graph.} A directed node-link diagram representing the LLM's internal task structure. Each node corresponds to a discrete subtask (\eg creating directories, validating input), and edges encode dependencies or data flow between them. This graph abstracts execution logic at the task level, without binding to specific code implementations (Fig.~\ref{fig:triple} right).

    \item \texttt{Mapping.} A cross-structure alignment that links each intent node to a corresponding node or subgraph in the Understanding Graph. This mapping ensures semantic consistency between user goals and generated code tasks, and supports graph-level operations such as task expansion, simplification, and selective editing based on user intent (Fig.~\ref{fig:triple} middle).
\end{itemize}

\subsubsection{SLM Fine-tuning with a Distillation Pipeline}
\label{sec:triple_pipeline}

To address the inefficiency caused by two-round LLM calls, we follow the common practice of fine-tuning existing SLMs for specific coding tasks. We aim to fine-tune an SLM that can (1) generate triples from prompts in a single step, bypassing the need for intermediate code generation, and (2) ensure that the triples extracted directly from prompts align closely with those produced by the two-stage extractor, which depends on intermediate code. \rv{The error propagation during multi-round interaction will be mitigated by the self-healing ability of the fine-tuned SLM and users' direct modification.}

\begin{figure}
    \centering
    \includegraphics[width=\linewidth]{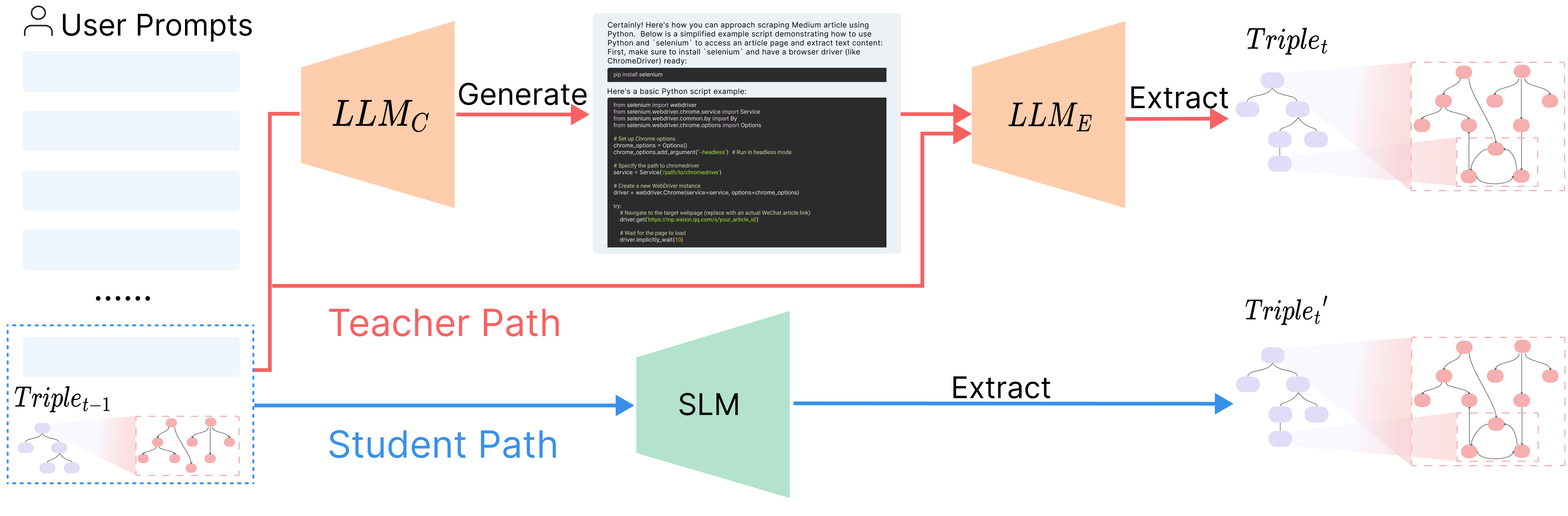}
    \caption{Triple Distillation Pipeline. It aligns the SLM in the student path with the two-stage extractor in the teacher path. The SLM can extract triples directly from prompts, bypassing intermediate code generation to speed up triple extraction.}
    \label{fig:distillation}
\end{figure}

We draw on knowledge distillation~\cite{KD}, a model compression technique that uses a teacher-student framework to transfer knowledge from a large, complex model (the teacher) to a smaller, more efficient model (the student). Knowledge distillation is well-suited for our goals because it enables the SLM (student) to replicate the outputs of the two-stage extractor (teacher) while being faster.
Specifically, our pipeline has a teacher path and a student path, utilizing three models (Fig.~\ref{fig:distillation}): a small language model (\(SLM\)), a conversational language model (\(LLM_c\)), and a triple extraction language model (\(LLM_e\)). For clarity, we define the following symbols: triples \(T_t\), prompts \(P_t\), and intermediate code \(C_t\), where \(t\) represents different rounds. 


\begin{figure*}
    \centering
    \includegraphics[width=\linewidth]{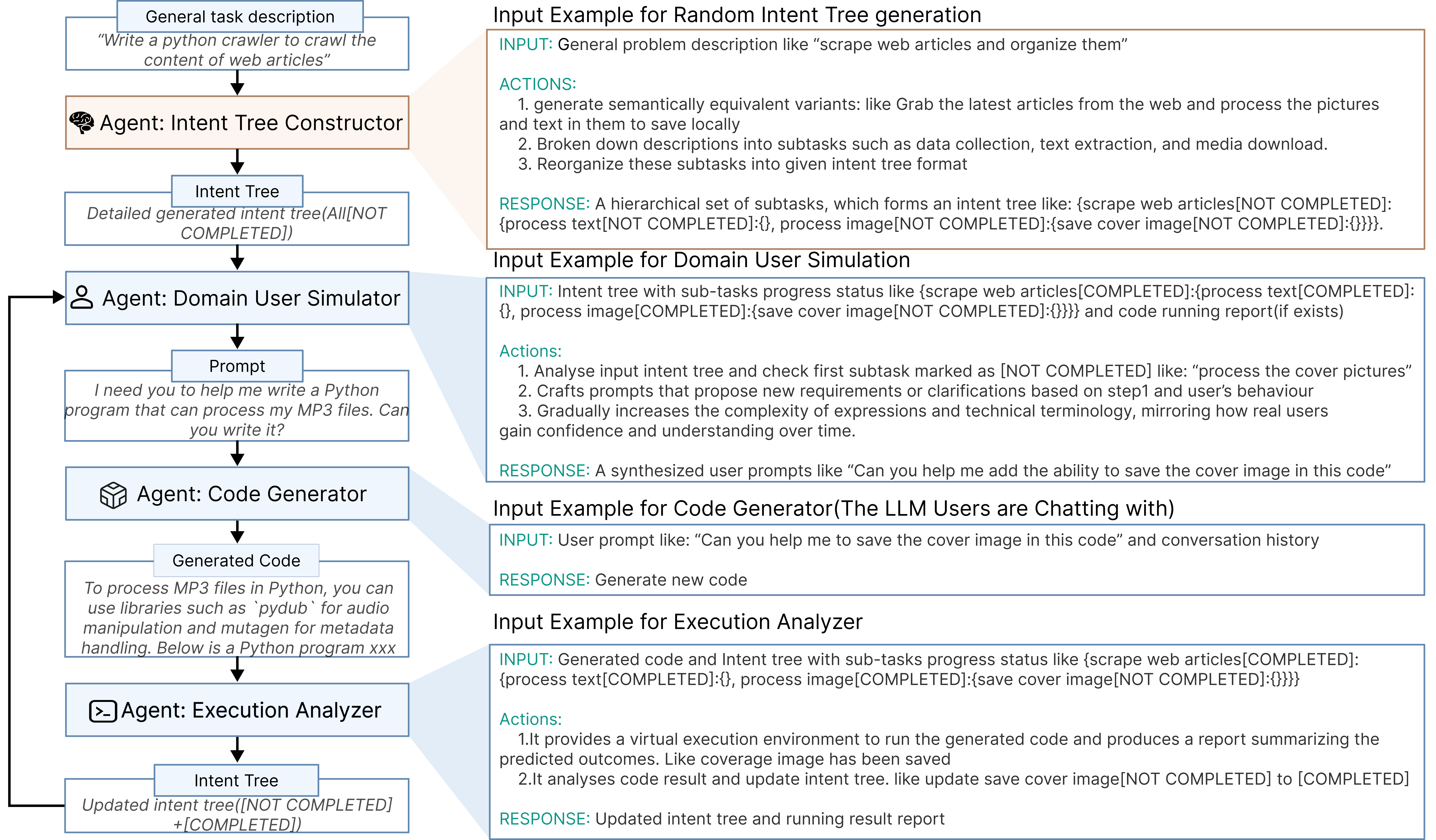}
    \caption{Multi-Agent Module Overview: This module involves four agents designed to interact with each other, simulating a domain user's experience of leveraging an LLM for code generation based on our findings on user behavior patterns.}
    \label{fig:playground}
\end{figure*}

\textbf{Teacher Path}.
We embed the two-stage extractor (Fig.~\ref{fig:probe}) into the teacher path. First, intermediate code is generated by \(LLM_c\), and then triples \(T_t\) are extracted using \(LLM_e\), based on the previous round's triples \(T_{t-1}\), the current round's prompt \(P_t\), and the intermediate code output \(C_t\). Specifically, we have $T_t = LLM_e(P_t, LLM_c(P_t), T_{t-1})$.
Here, \(LLM_e\) can be any sufficiently powerful model.

\textbf{Student Path}.
To generate more accurate triples without relying on the code produced by \(LLM_c\), it is necessary to construct a knowledge base that connects the beginning and the end of the teacher path and transfer this knowledge to an SLM. To achieve this, we design the student path and train the \(SLM\). To preserve the original generalization ability of the SLM, we incorporate a LoRA \cite{hu2022lora} adapter, keeping the original parameters of the SLM fixed while tuning only the adapter's parameters. The SLM then directly generates the current round's triples \(\hat{T_t}\) based on the current prompt \(P_t\) and the previous round's triples \(T_{t-1}\), expressed as $T_t = SLM(P_t, T_{t-1})$.

\textbf{Alignment}.
We use the mean squared error (MSE), $\text{MSE} = \frac{1}{n} \sum_{i=1}^{n} (\hat{T_t}_i - {T_t}_i)^2$, to align \(\hat{T_t}\) and \(T_t\) (the outputs of the student path and teacher path, respectively) across all output tokens. Here, \(i\) refers to the token index. In this way, the SLM can bridge the beginning (\(P_t, T_{t-1}\)) and end (\(T_t\)) of the teacher path without generating intermediate code.

\subsubsection{Dataset Generation with a Multi-Agent Module}
\label{sec:multiagent}
The distillation pipeline relies on user prompts as input. However, collecting prompts directly from real users is costly and time-consuming. To address this challenge, we introduce a multi-agent module designed to efficiently synthesize prompts. Our formative findings reveal that domain users typically initiate the problem-solving process with an LLM by crafting prompts with a general problem description. They then receive code generated by an LLM, execute it directly without closely reading the code, and refine their prompts for subsequent interactions based on where the results do not match their expectations. 
To simulate this process, we design the multi-agent system with four specialized agents. Below, we outline the roles of these agents and how they collaborate to generate prompts effectively, with an example shown in Fig.~\ref{fig:playground}.

\textbf{Agent Design.}
All four agents are built on LLMs, using well-designed prompts to generate realistic responses:
\begin{itemize}[left=0pt]
    \item \textit{Intent Tree Constructor}. This agent takes a general problem description as input and then decomposes it into a hierarchical set of subtasks, which forms an intent tree. For example, a description of \q{scrape web articles and organize them} might be broken down into subtasks such as data collection, text extraction, and media download. For each description, we ensure the agent can generate semantically equivalent variants in different expressions, accommodating the varying ways users might phrase their tasks.
    
    \item \textit{Code Generator}. This agent takes synthetic prompts as input and outputs generated code along with comments. It functions the same as an LLM (\eg ChatGPT) that a domain user interacts with in real-world scenarios. 
    
    \item \textit{Execution Analyzer.} This agent simulates a user executing code and comparing the results with their expectations. Specifically, it provides a virtual execution environment to run the generated code and produces a report summarizing the predicted outcomes, file changes, and any errors encountered. Additionally, it analyzes the execution results against the subtasks in the intent tree, updating the state of each subtask as either [COMPLETED] or [NOT COMPLETED].
    
    \item \textit{Domain User Simulator}. This agent takes an intent tree with states as input and synthesizes user prompts. Specifically, it crafts prompts that propose new requirements or clarifications based on the first subtask marked as [NOT COMPLETED], simulating how a real user's intent is often shaped by the first unexpected result. The agent is instructed to craft prompts in the style of a domain user without coding expertise, using emotional markers to make user reactions more realistic. As the conversation progresses, the agent gradually increases the complexity of expressions and technical terminology, mirroring how real users gain confidence and understanding over time.    
\end{itemize}


\textbf{Prompt Generation with Agents.} 
With the four agents, each execution of the following process generates a multi-round prompt history for a single problem:

\begin{itemize}
    \item \textit{Initialization}: The process begins with a general task description, which the \textit{Intent Tree Constructor} uses to generate an intent tree. This tree serves as the foundation for identifying and tracking subtasks throughout the process. 
    \item \textit{Collaboration Loop}: The system enters an iterative loop of prompt generation, code generation, and state updates. Specifically, the \textit{Domain User Simulator} identifies subtasks marked as [NOT COMPLETED] and crafts refined prompts. These prompts are passed to the \textit{Code Generator}, which produces executable code and detailed explanations. The \textit{Execution Analyzer} then runs the code, evaluates the results, and updates the task status within the intent tree. This iterative loop continues, with each agent collaboratively refining prompts and advancing task completion. 
    \item \textit{Termination}: The process ends when all subtasks are [COMPLETED], or if no progress occurs over five consecutive dialogue rounds. This ensures the workflow is both goal-oriented and time-efficient.
\end{itemize}

\subsection{Graph Simplifier}
\label{sec:Simplification}

\begin{figure}
    \centering
    \includegraphics[width=\linewidth]{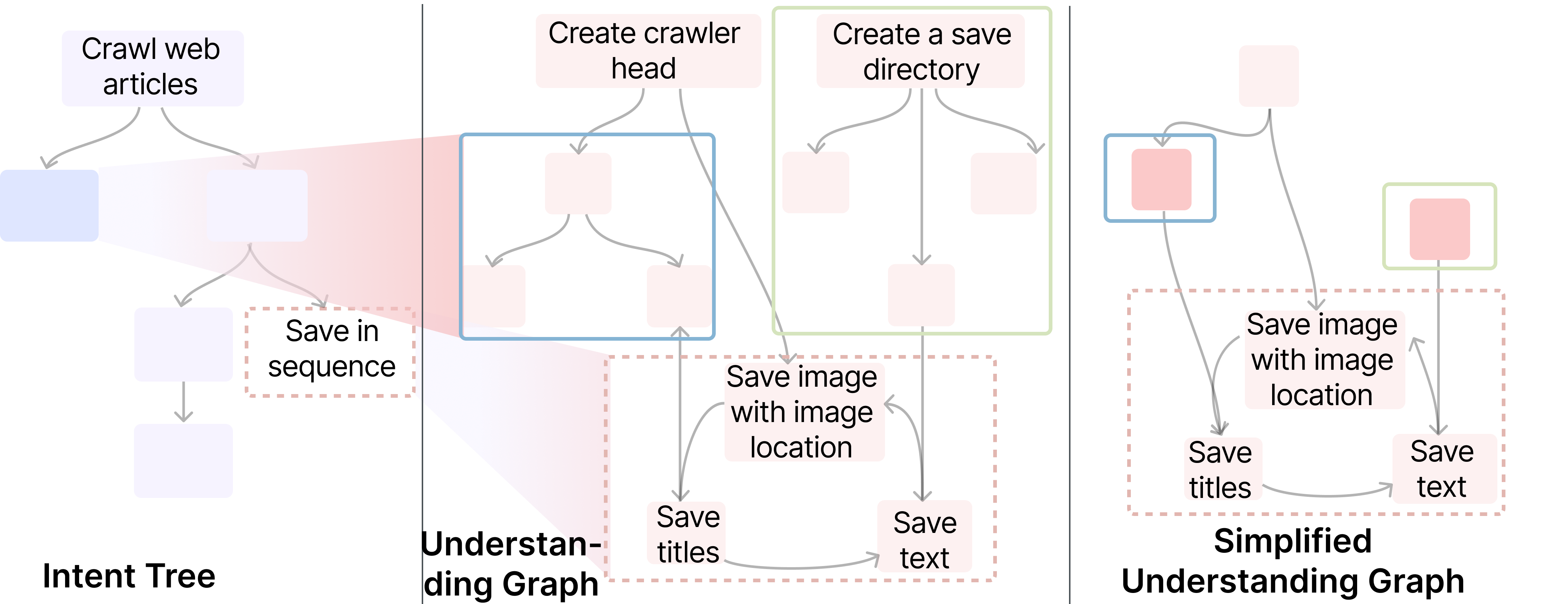}
    \caption{Intent-aware graph simplification algorithm. The left figure illustrates an intent tree, where each node corresponds to a sub-understanding graph. During the simplification process, nodes that are mapped to changes in the intent tree are directly transferred to the simplified graph (\ie~red dashed box). Meanwhile, parts mapped to unchanged nodes are recursively collapsed or zoomed out (\ie~blue and green boxes). 
    }
    \label{fig:graph_sim}
    
\end{figure}

Since users often lose track of how the complex LLM understanding graph evolves as their intent changes, they require the graph to be simplified in alignment with their updated intent (\textbf{DC3}). To address this, we propose an intent-aware graph simplification algorithm that highlights nodes directly related to intent changes and collapses other nodes (Fig.~\ref{fig:graph_sim}). It operates in two stages: intent tracking and graph simplification.

\textbf{Intent tracking}. The algorithm explicitly tracks user intent changes across multiple dialogue rounds. To achieve this, we construct a Nondeterministic Finite Automaton that expresses and automatically updates user intents. Initially, the algorithm analyzes the user's input, extracts the overall goal, and identifies specific subtasks or requirements, organizing this information into an intent tree. As the dialogue progresses, the Intent Tree is updated to reflect the user's latest input. These updates may involve refining existing intents, introducing new intents, merging similar intents, adjusting relationships between parent and child nodes, or confirming that no changes are necessary. After every update, the algorithm synchronizes the Intent Tree with the understanding graph, ensuring that each intent node corresponds to a subgraph within the task graph. This synchronization provides a clear mapping between user intentions and the graph structure, enabling the targeted simplification of the graph in the next stage.

\textbf{Graph simplification}. 
The algorithm employs recursive topological reduction of the hierarchical intent tree. Given a focus node set $F \subseteq \mathcal{V}_T$ ($\mathcal{V}_T$ being the set of intent tree nodes), the algorithm recursively traverses from the second-layer nodes $\{v^{(2)}_i\}$: for any node $v$, if its sub-tree $\mathcal{T}(v)$ satisfies $F \cap \mathcal{T}(v) = \emptyset$, the corresponding subgraph $G_v \subseteq G$ is collapsed into a supernode $u$, establishing a mapping $\phi: V(G_v) \to u$; otherwise, it recursively checks the direct child nodes $\{c_j\}$ of $\mathcal{T}(v)$, repeating the above judgment for each $c_j$. Ultimately, all subgraph structures containing members of $F$ are preserved, unrelated branches are merged, and the edge set across subgraphs $\bigcup \{(s',t') | s'=\phi(s), t'=\phi(t), (s,t) \in E\}$ is reconstructed through $\phi$, where $\phi(x)=x$ if and only if $x \notin \bigcup V(G_{v_i})$ ($v_i$ being the merged nodes). The corresponding process is shown below (Fig.~\ref{fig:graph_sim}), ensuring graph simplification while balancing global awareness and cognitive load.

\begin{table}[t]
\setlength{\tabcolsep}{4.15pt}
\caption{Comparison of three fine-tuned SLMs and their zero-shot counterparts. Higher values indicate greater similarity to the ground truth produced by the two-stage extractor.}
\label{tab:acccomp}
\begin{tabular}{ccccccc}
\toprule
 & \multicolumn{2}{c}{\textbf{Qwen 1.5B}} & \multicolumn{2}{c}{\textbf{Qwen 7B}} & \multicolumn{2}{c}{\textbf{LLaMa 8B}} \\ 
\cmidrule(lr){2-3} \cmidrule(lr){4-5} \cmidrule(lr){6-7}
\textbf{Metric} &
\parbox[t]{0.8cm}{\centering Zero-shot} & 
\parbox[t]{0.8cm}{\centering Fine-tuned} & 
\parbox[t]{0.8cm}{\centering Zero-shot} & 
\parbox[t]{0.8cm}{\centering Fine-tuned} & 
\parbox[t]{0.8cm}{\centering Zero-shot} & 
\parbox[t]{0.8cm}{\centering Fine-tuned} \\ 
\midrule
ROUGE-1 & 0.8503 & \textbf{0.8946} & 0.8590 & \textbf{0.9099} & 0.8621 & \textbf{0.9274} \\
ROUGE-2 & 0.6941 & \textbf{0.7884} & 0.7259 & \textbf{0.8209} & 0.7277 & \textbf{0.8545} \\
ROUGE-L & 0.7872 & \textbf{0.8644} & 0.8203 & \textbf{0.8885} & 0.8214 & \textbf{0.9126} \\
BLEU    & 0.8598 & \textbf{0.9214} & 0.8747 & \textbf{0.9340} & 0.8741 & \textbf{0.9434} \\ 
\bottomrule
\end{tabular}
\end{table}

\begin{figure}[t]
    \centering
    \includegraphics[width=\linewidth]{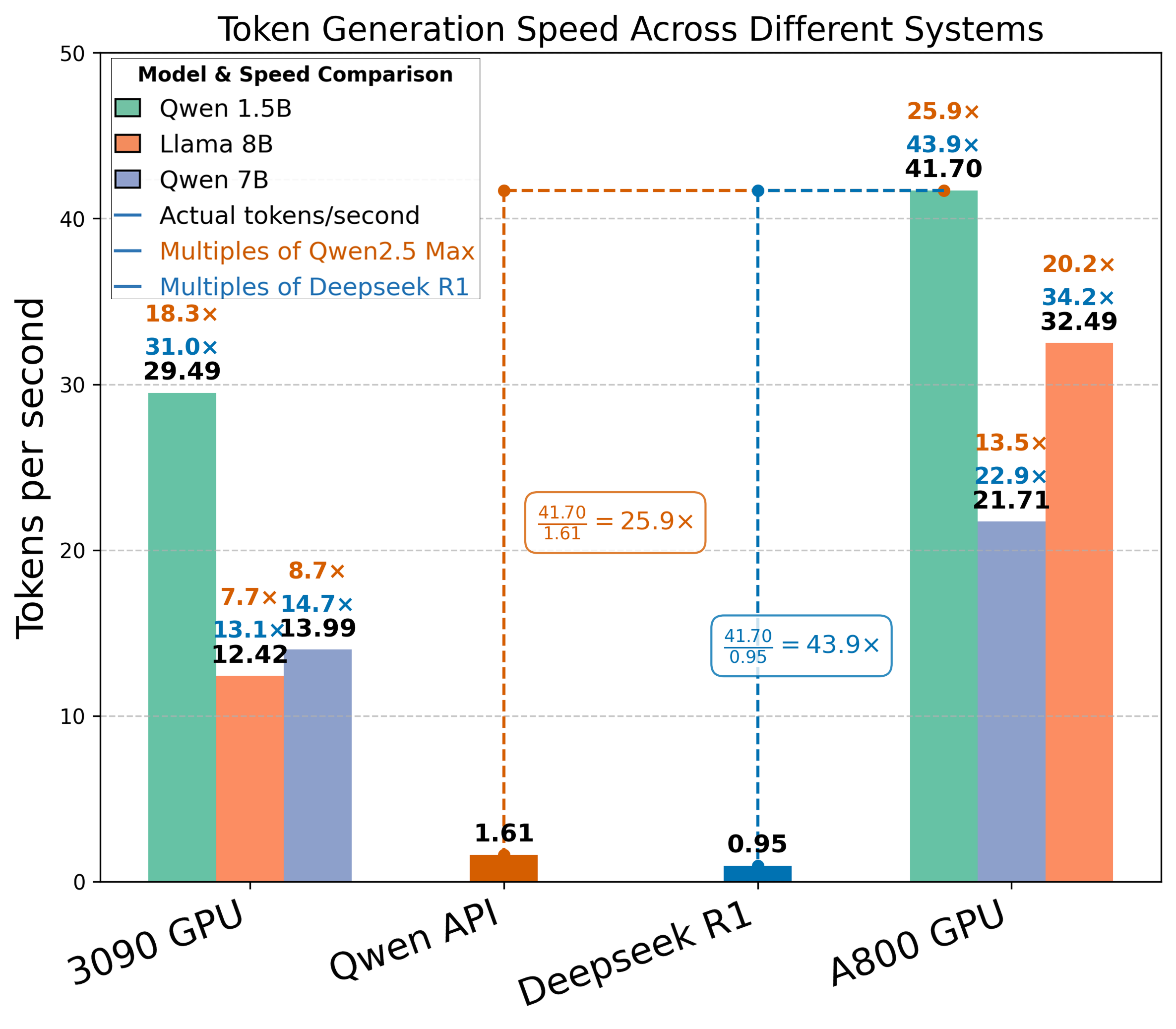}
    \caption{Efficiency gains in SLM inference speed over the two-stage extractor across different hardware settings. Higher values indicate faster extraction speed.}
    \label{fig:tokenpersecond}
\end{figure}

\setlength{\parindent}{8pt}

\section{Technical Evaluation}

This section evaluates how well our distillation pipeline (Sec.~\ref{sec:triple_pipeline}) transfers knowledge from the two-stage extractor with two-round LLM calls to an SLM (Fig.~\ref{fig:probe}) for triples extraction. The pipeline is designed to fine-tune the SLM to mimic the behavior of the two-stage extractor while being more lightweight. Therefore, we assess the SLM from two aspects: (1) \textit{similarity}, which measures how closely the triples extracted by the SLM align with those extracted by the two-stage extractor, and (2) \textit{efficiency}, which evaluates the speed of the SLM compared to the two-stage extractor. 

\subsection{Similarity Between SLM Outputs and the Two-Stage Extractor Outputs}
\label{subsec:similarity}

In this evaluation, we compare the outputs of the fine-tuned SLMs with their zero-shot counterparts (acting as baselines), using the outputs of the two-stage extractor as the ground truth.

\textbf{Metrics.}
To measure similarity, we employ two widely used metrics in nature language processing: ROUGE~\cite{rouge} and BLEU~\cite{bleu}. ROUGE is a recall-based metric that evaluates how much of the reference text (i.e., the 2-stage method’s outputs) is covered in the target text (i.e., the SLM’s outputs). Specifically, ROUGE-1 measures overlaps of single words, ROUGE-2 measures overlaps of two consecutive words, and ROUGE-L evaluates the longest common subsequence between the texts. BLEU, on the other hand, is a precision-based metric that evaluates how well the target text matches the reference while penalizing irrelevant content. ROUGE and BLEU provide a balanced evaluation of content coverage and precision, assessing how well the SLMs replicate the LLM.

\textbf{Baselines.}
We considered three pre-trained SLMs in our experiments: Qwen-2.5 1.5B, Qwen-2.5 7B, and Llama 8B. They served both as zero-shot baselines and as the foundation for the student models trained during the distillation process.

\textbf{Experiments.}
Our experiments consisted of three phases: SLM fine-tuning, inference, and comparison. During SLM fine-tuning, we placed each pre-trained SLM in the student path and fine-tuned each of them on a server with three NVIDIA 3090 GPUs. This process produced three distilled SLMs corresponding to Qwen-2.5 1.5B, Qwen-2.5 7B, and Llama 8B, respectively. Zero-shot baselines used the pre-trained models without fine-tuning, enabling a comparison of improvements from the distillation process. In the inference phase, fine-tuned and zero-shot SLMs generated triples for a testing dataset of 40 samples per epoch, producing one triple per sample. The teacher path’s two-stage extractor also generated one triple per sample, serving as ground truth. Lastly, we compared the triples from fine-tuned and zero-shot SLMs to the ground truths using ROUGE and BLEU, averaging the values across the 40 samples in the final epoch.

\textbf{Result analysis.} 
The results in Tab.~\ref{tab:acccomp} demonstrate the effectiveness of our distillation pipeline, with fine-tuned SLMs outperforming their zero-shot counterparts across all metrics. Specifically, these results highlight the pipeline’s ability to enhance both content coverage and precision, achieving alignment with the two-stage extractor's outputs at over 90\% similarity in most metrics. Larger models, such as LLaMa 8B, show greater improvements. While slight performance gaps remain due to the inherent randomness in language generation, they do not hinder the overall alignment and effectiveness of the fine-tuned models.

\subsection{Efficiency Gains in SLM Inference Speed}

In this evaluation, we compare the inference efficiency of fine-tuned SLMs to the two-stage extractor for triple extraction.

\textbf{Metrics.}
We calculate \textit{valid tokens per second}, which measures the number of tokens generated per second that correspond directly to triples. Unlike the commonly used token per second (\ie~$Overtime\ Tokens/Overall\ Time$), this metric excludes irrelevant intermediate tokens, offering a more focused comparison between the fine-tuned SLMs and the two-stage extractor.

\textbf{Baselines.}
The two-stage extractor serves as the baseline and we consider two variants of its implementation. Specifically, the two-stage extractor involves two LLM calls: the first LLM, Deepseek R1~\cite{deepseekr1}, is used to generate code, while the second LLM extracts triples based on the generated code. For the second LLM, we evaluate two options: Qwen-Max~\cite{Qwen}, an SOTA general LLM, and DeepSeek R1~\cite{deepseekr1}, an SOTA reasoning LLM.

\textbf{Experiments.}
We evaluated the efficiency of fine-tuned SLMs and the two-stage extractor variants on two hardware platforms: an NVIDIA 3090 server and an NVIDIA A800 server. The 3090 server simulates a setup for individual users, while the A800 server reflects a deployment used in industry settings. Each configuration was tested on 40 user prompts. For each prompt, we calculated valid tokens per second based on the total processing time and the number of valid tokens generated. The results were averaged across all prompts. 

\textbf{Result analysis.}
The results in Fig.~\ref{fig:tokenpersecond} highlight the significant efficiency gains achieved by the fine-tuned SLMs. The distilled SLMs consistently outperformed the two-stage extractor variants at both server settings. For example, using Llama 8B, the A800 server achieved a 22.9× speed improvement compared to the two-stage extractor with Deepseek R1, and the same model on the 3090 server achieved a 13.1× speedup. These findings indicate that our proposed method is advantageous for both individual and industrial applications.

\begin{table}[t!]
\caption{Participant demographics in the user study.}
\label{tab: userstudy}
\renewcommand\arraystretch{1}
\setlength{\tabcolsep}{2mm}{
\begin{tabular}{lcccc}
\toprule
ID  & Gender & Age & Education & Domain Expertise \\ \midrule
P1  & M      & 24  & Ph.D      & Theoretical Mathematics \\
P2  & M      & 23  & Ph.D      & Operations Research \\
P3  & F      & 23  & M.S.      & Art \\
P4  & F      & 23  & M.S.      & Design \\
P5  & M      & 26  & Ph.D      & Civil Engineering \\
P6  & M      & 24  & Ph.D      & Economics \\
P7  & F      & 25  & Ph.D      & Linguistics \\
P8  & F      & 24  & M.S.      & Design \\
P9  & M      & 23  & M.S.      & Finance \\
P10 & F      & 24  & Ph.D      & Biology \\
P11 & M      & 28  & M.S.      & Architecture \\
P12 & F      & 23  & M.S.      & Education \\ \bottomrule
\end{tabular}}
\end{table}

\begin{figure*}[t]
    \centering
    \includegraphics[width=\linewidth]{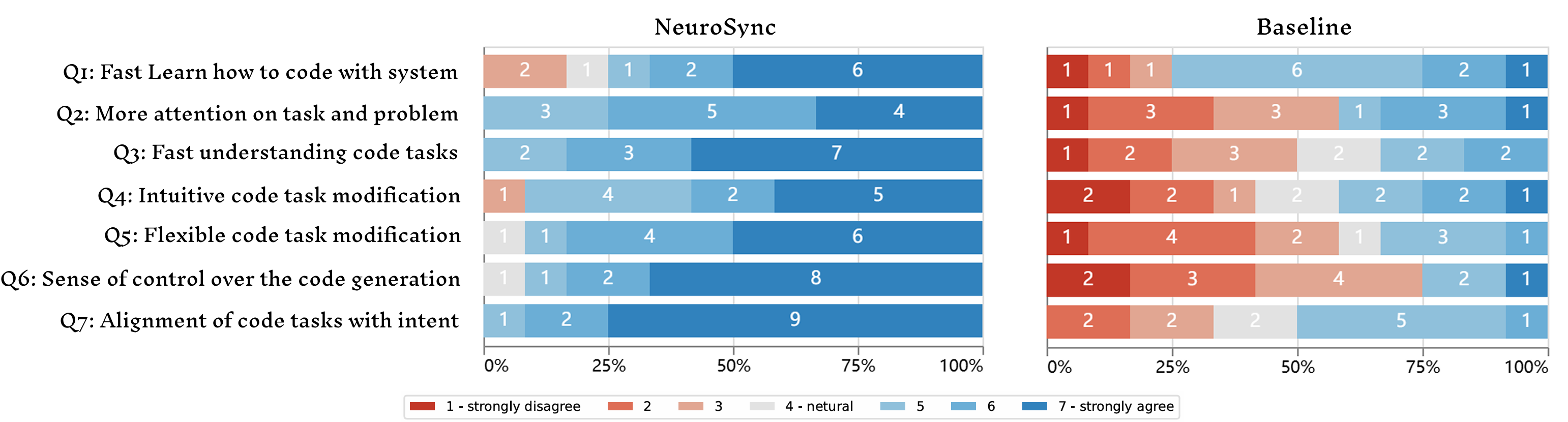}
    \caption{User ratings on the baseline and \tool with a 7-point Likert scale.}
    \label{fig:sus}
\end{figure*}

\section{Controlled User Study}

To evaluate the usability and effectiveness of \tool in supporting 
conversational LLM-based coding, we conducted a controlled study with 12 domain users with limited programming experience. The study addressed two primary research questions:

\vspace{0.5\baselineskip}
\noindent\textbf{RQ1:} Can \tool improve domain users' ability to effectively perform conversation-based coding with LLMs?
\begin{itemize}[left=22pt]
    \item[\textit{RQ1.1:}] Does the graph-based representation facilitate task comprehension and reduce barriers to real-time coding?
    \item[\textit{RQ1.2:}] Can \tool support more precise, controlled modifications to users' task-level understanding?
    \item[\textit{RQ1.3:}] Does \tool help domain users accomplish intent-aligned code generation more efficiently,  with fewer interactions?
\end{itemize}

\noindent\textbf{RQ2:} How does \tool affect user perceptions and behaviors during LLM-based coding interactions?
\begin{itemize}[left=22pt]
    \item[\textit{RQ2.1:}] How does the system shift users' mental models regarding coding with LLMs?
    \item[\textit{RQ2.2:}] How does the interaction paradigm influence user behaviors in multi-turn coding?
    \item[\textit{RQ2.3:}] What additional or unforeseen impacts does using \tool introduce?
\end{itemize}

\subsection{Methods}

\subsubsection{\textbf{Participants.}} 
We recruited 12 postgraduate students (six males, six females) aged 23 to 28 years ($M=24.17$, $SD=1.53$), from diverse domains such as art and design, linguistics, education, economic, finance, and architecture. Participants self-rated themselves as having general experience within their own domain ($M=3.58$, $SD=0.51$), but limited familiarity with coding ({$M=2$, $SD=0.85$}). They reported prior experience with LLM-powered chatbots ($M=4.16$, $SD=0.73$), based on a 5-point Likert scale (1 = lowest, 5 = highest). Detailed demographic information is provided in Tab.~\ref{tab: userstudy}.


\subsubsection{\textbf{Tasks.}}
Participants completed two programming tasks: a web crawler task and an audio processing task. In the \textit{web crawler task}, they implemented a crawler that retrieved content from a specified WeChat article URL. 
In the \textit{audio processing task}, participants developed a Python script that converted MP3 audio into text and extracted keywords relevant to sentiment analysis. 



\subsubsection{\textbf{Baseline.}} 
We used a simplified version of \tool, referred to as the Baseline, which excluded graphical representations while preserving conversational features (\eg ChatGPT-style prompting). The baseline included standard Python syntax highlighting and detailed inline code comments. Both experimental conditions maintained identical user interface styles to control for differences in visual aesthetics. \rv{The read-only task graph was not considered a baseline because feedforward and editing were designed as an integrated mechanism to address misalignment; evaluating them separately would not reflect their intended use.}

\subsubsection{\textbf{Apparatus.}} All participants completed the tasks on desktop environments, each connected via SSH to a GPU server equipped with an NVIDIA A800.

\subsubsection{\textbf{Procedure.}}
We employed a counterbalanced within-subjects design. Participants were divided into two clusters, each performing both tasks, one task using \toole, the other using the baseline. Task order and condition assignments were balanced within clusters. 
Each session comprised an introduction (5 mins), two task sessions (10–45 mins each), post-task questionnaires (10 mins each), and a final semi-structured interview (15–25 mins). Interviews were audio-recorded, and each participant was compensated \$12/hour (approximately 1.5 hours per session).
To ensure the tasks remained exploratory and truly started from scratch, we designed the task session by distributing task information and completion criteria across two complementary communication channels, each serving a different purpose. Participants first received a written task description outlining high-level goals, followed by oral delivery of more detailed background context. This separation was intended to promote diverse problem-solving approaches and avoid uniform strategies, such as copying all requirements directly into the LLM. After task presentation, participants were asked to articulate their understanding to confirm comprehension before proceeding independently. Once they believed they had completed the task, the facilitator reviewed their outcome against general completion criteria and provided feedback indicating whether the task was complete or needed further refinement.





\subsubsection{\textbf{Measurements.}} 
We evaluated both systems across three aspects: {System Usability}, {Cognitive Load}, and {Coding Efficiency}. \textit{System Usability} was measured using a custom 7-point Likert questionnaire that evaluated learnability, code comprehension, task modification accuracy, and alignment. \textit{Cognitive Load} was assessed using the NASA-TLX, which included a measure of perceived cooperation to capture subjective mental workload differences between conditions. \textit{Coding Efficiency} was evaluated through task completion times, durations of task-focused thinking and manipulation (excluding waiting time for LLM inference), and the number of LLM queries made during the tasks.


\begin{table}[t!]
\centering
\caption{
Comparison of mean scores in 7 questions (Fig.~\ref{fig:sus}) between the baseline and \tool with statistical analysis. \textit{M.} denotes the mean score for each system, \textit{Diff.} indicates the difference, and \textit{t} and \textit{p} report the paired t-test results.
}
\label{tab:sus_comparison}
\setlength{\tabcolsep}{6pt}
\begin{tabular}{cccccc}
\toprule
\textbf{Dim.} & \textbf{$M.(Baseline)$} & \textbf{$M.(Ours)$} & \textbf{Diff.} & \textbf{t} & \textbf{p} \\
\midrule
Q1 & 4.58 & 5.75 & 1.17 & 2.18 & .05150\\
Q2 & 3.83 & 6.08 & 2.25 & 3.28 & .00738\\
Q3 & 3.67 & 6.42 & 2.75 & 5.25 & .00027\\
Q4 & 3.83 & 5.83 & 2.00 & 3.32 & .00687\\
Q5 & 3.33 & 6.25 & 2.92 & 5.24 & .00028\\
Q6 & 3.08 & 6.42 & 3.33 & 5.61 & .00016\\
Q7 & 4.08 & 6.67 & 2.58 & 6.49 & .00004\\
\bottomrule
\end{tabular}
\end{table}

\subsubsection{\textbf{Analysis.}} 
Throughout the study, we collected audio recordings, interaction logs, and questionnaire data. \rv{We used an open-coding approach~\cite{opencode} for data analysis.} For quantitative measures that met assumptions of normality and homogeneity of variance, paired t-tests were employed to assess statistical significance. Subjective ratings were evaluated using the Wilcoxon signed-rank test. Audio recordings were transcribed and categorized based on the research questions. The results are presented according to this analytical approach.

\subsection{Quantitative Results}

To evaluate \tool against the Baseline, we assessed performance across three dimensions: \textit{system usability for coding}, \textit{cognitive load}, and \textit{coding efficiency}. 

\subsubsection{\textbf{System Usability for Coding (RQ1)}}

To assess usability, we administered a questionnaire (Q1–Q7) using a 7-point Likert scale, covering learnability (RQ1.1), code comprehension (RQ1.1), task modification (RQ1.2), and misalignment reduction (RQ1.3). As shown in Fig.~\ref{fig:sus} and Tab.~\ref{tab:sus_comparison}, \tool consistently outperformed the baseline across all dimensions:

\textbf{Lower Learning Threshold.} \tool showed an improvement in learnability (Q1; Baseline: 4.58 vs. \toole: 5.75; $p = .051$). While the addition of graph-based interactions introduced new functionality, users found the system easier to learn.
    
\textbf{Improved Task Comprehension.} Participants rated the graph representation as highly intuitive for understanding code logic (Q3; mean diff = 2.75, $p < .001$) and focusing on task structure (Q2; mean diff = 2.25, $p < .001$), enabling direct mapping between intent and generated functionality.

\textbf{Enhanced Control and Modification.} \tool significantly enhanced users' ability to identify, modify, and refine code tasks (Q4–Q6; all $p < .01$). The explicit task-level editing mechanism allowed users to bypass abstract prompt tuning, reducing cognitive overload and frustration.

\textbf{Reduced Intent–Code Misalignment.} Participants reported fewer instances of code diverging from their original intentions when using \tool (Q7; mean diff = 2.58, $p < .001$), validating the effectiveness of task-level alignment.

\begin{figure}[t]
    \centering
    \includegraphics[width=\linewidth]{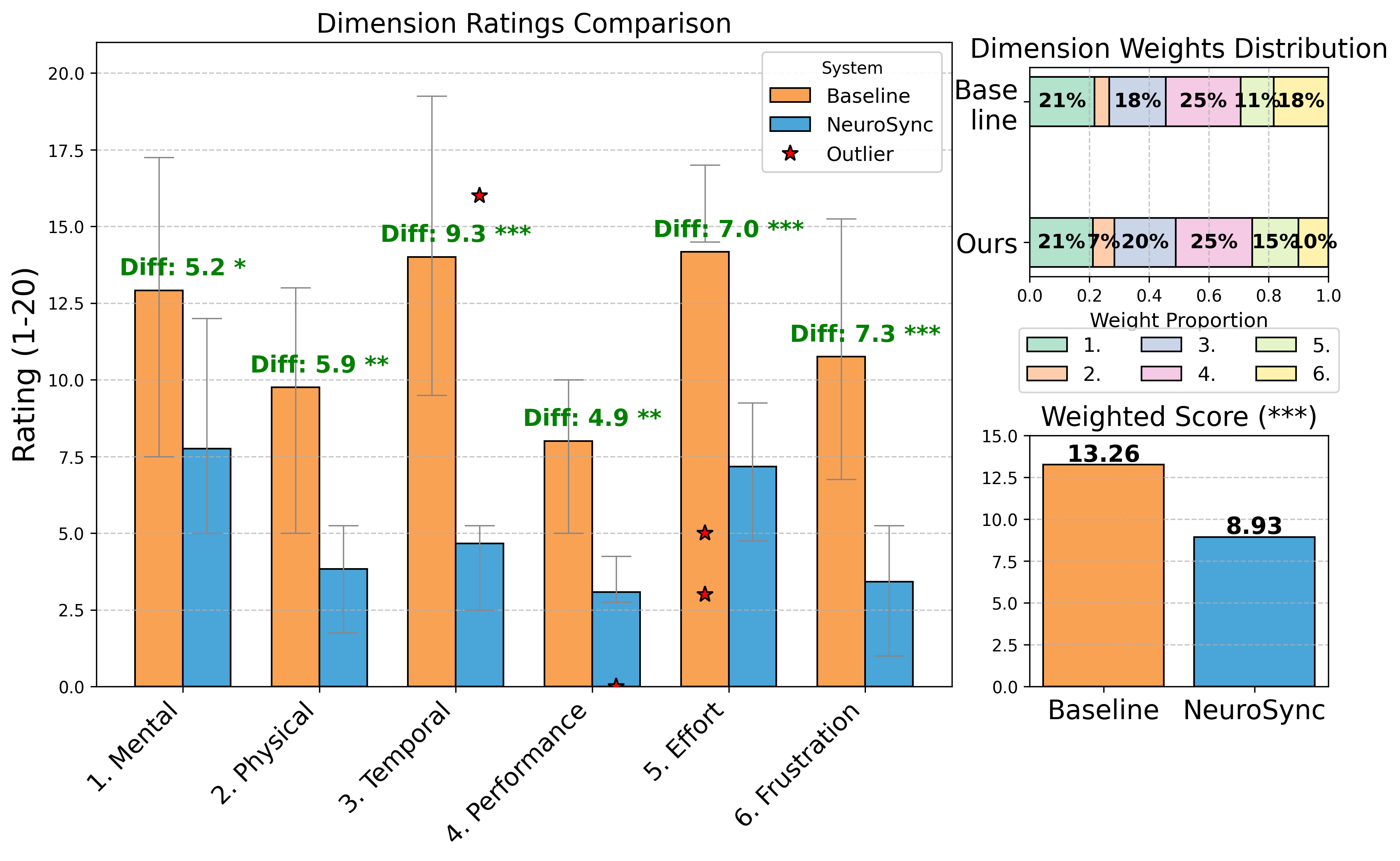}
    \caption{User ratings on the baseline and \tool using NASA-TLX. \rv{Dimension weight shows each dimension’s relative importance to the overall workload.}}
    \label{fig:tlx}
\end{figure}

\subsubsection{\textbf{Cognitive Load (RQ2.1)}}

We used NASA-TLX to evaluate perceived mental workload after completing tasks with each system. As shown in Fig.~\ref{fig:tlx}, \tool led to significantly lower cognitive load across all six dimensions:

\textbf{Overall Load Reduction.} Participants experienced lower total workload with \tool (Baseline: 13.26 vs. \toole: 8.93; $p < .001$), especially in task time demand (D3; diff = 9.3, $p < .001$) and frustration (D6; diff = 7.3, $p < .001$), with frustration levels dropping from 18\% to 10\%.

\textbf{Performance Not a Bottleneck.} Although performance scores (D4) improved with \tool (diff = 4.9, $p < .01$), this dimension showed the smallest margin. This suggests that LLMs already met baseline task requirements, and the benefit of \tool was in making their output more controllable and understandable.

\textbf{Shifted Cognitive Effort.} Mental Demand (D1) and Effort (D5) were slightly higher than other \tool dimensions (D1 = 7.75; D5 = 7.16), though still significantly lower than Baseline. This reflects the tradeoff: while \tool reduces effort in code understanding, it introduces new cognitive demands in interacting with the graph.

\subsubsection{\textbf{Coding Efficiency (RQ2.2)}}

\begin{figure}[t]
    \centering
    \includegraphics[width=\linewidth]{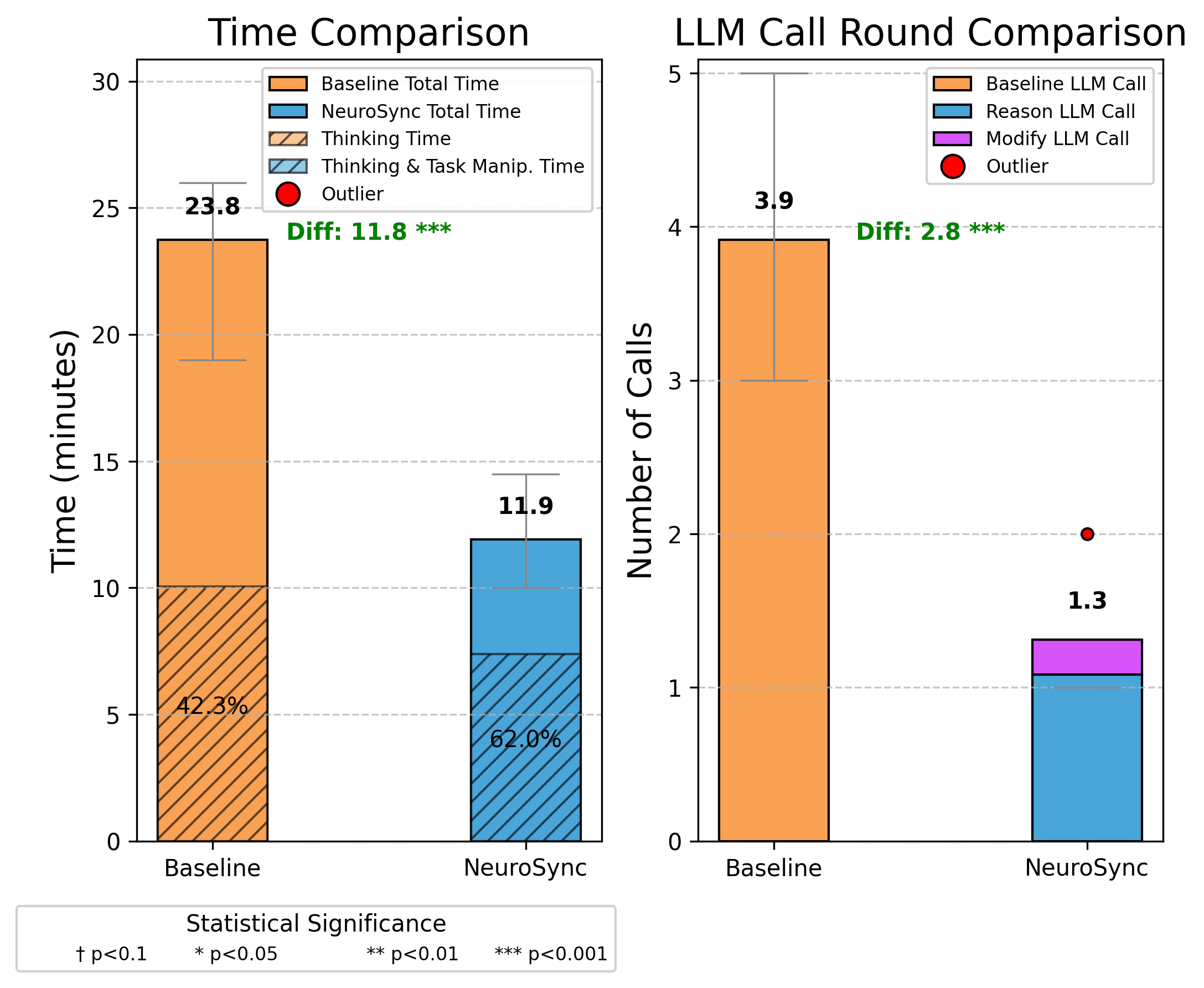}
    \caption{Quantitative results of baseline and \tool on time consumption and LLM call rounds.}
    \label{fig:runtime}
\end{figure}

We recorded task duration, user thinking/manipulation time (excluding LLM wait time), and total LLM calls per task. As shown in Fig.~\ref{fig:runtime}, key findings included:

\textbf{Faster Task Completion.} Participants using \tool completed tasks significantly faster (23.8 mins vs. 13.9 mins; $p < .001$), with fewer LLM calls (3.9 vs. 1.3; $p < .001$). This demonstrates that task-level editing reduced the need for iterative prompt correction.

\textbf{Increased Task Focus.} With \toole, users spent more time on task reasoning (62\% vs. 42.3\%) rather than interpreting or rewriting prompts. The externalization of task structure helped them focus on problem-solving rather than system communication.

\subsection{Qualitative Insights}
Through post-task interviews, participants expressed generally positive feedback on the usability and effectiveness of \toole. 

\textbf{Improved Task Understanding.}
One of the most recognized advantages was its ability to assist users in understanding the structure and logic of code through the pre-generated task graph (P1–P7, P9–P10, P12). Compared with directly reading raw code, the graph provided a clearer and more intuitive overview of task flow, especially for users with limited programming experience. As P3 mentioned, \textit{``I didn’t need to understand every line of code—I just looked at the flow and knew what was going on.''} This visual representation enabled participants to enter the problem-solving process more efficiently and reduced their reliance on reading and interpreting code syntax line by line.

\textbf{Reduced Programming Barrier.}
The graph-based interaction was considered effective in lowering the entry barrier for programming tasks. Several participants (P1, P3–P5, P8–P9, P12) commented that they could express their ideas more clearly through structured tasks in the graph, rather than struggling to describe them precisely using natural language prompts. For example, P4 stated, \textit{``The graph helped me break down what I wanted into manageable parts—without thinking about how to write it in code.''} This form of externalized task structure encouraged users to reflect on and refine their intentions more systematically, especially when handling multi-step tasks.

\textbf{More Accurate Modifications.}
Participants (P3–P9, P12) also emphasized the benefits of the graph-editing mechanism in improving task modification. Compared with traditional prompt-based interaction, direct manipulation of task nodes allowed for more accurate and targeted adjustments. As P6 said, \textit{``Instead of rewriting everything, I just fixed the node that was wrong and got what I wanted.''} Moreover, when users were uncertain about how to modify the graph manually, the natural language-based modification interface provided a convenient alternative (P2–P3, P7–P8, P10). P2 noted, \textit{``Sometimes I didn’t know how to change the graph directly, so I just typed what I wanted, and it worked.''} These complementary interaction modalities enhanced users' control over task editing.

\textbf{Fewer Dialog Turns.}
Furthermore, many participants (P2–P3, P5, P7–P10) reported that the system effectively reduced the number of interaction rounds required to align code with their intentions. The holistic representation of tasks, along with the ability to directly revise sub-tasks, allowed users to express and adjust their goals more clearly. As P8 observed, \textit{``Normally it takes me five tries to get it right. With this, I got most of it on the first go.''} This improvement in efficiency was particularly appreciated by users who had prior experience with LLM-based tools.

\textbf{Effective Graph Simplification.}
The graph simplification mechanism was also highly valued by all participants. The integration of the intent tree with the simplified task graph enabled users to efficiently identify key task components and understand overall logic at both macro and micro levels. P9 commented, \textit{``It’s like zooming out and zooming in at the same time. I could see the big picture and the small details without getting lost.''} The highlight feature, which linked simplified and detailed views, further facilitated focused editing and task tracing during multi-round interactions.

\rv{\textbf{Different Interaction Patterns.} The externalized LLM understanding shifted the participants’ interaction patterns in two ways. The first was misalignment resolution. With the baseline, participants often addressed only partial misalignments and introduced new ones due to incomplete reviews of code and prompts. In contrast, \toole’s task graph allowed them to detect and resolve all misalignments in a single pass by directly modifying relevant nodes, improving their understanding of the problem-solving process. The second was the timing of instruction. With the baseline, participants issued instructions reactively after each code generation round, resulting in more iterations. \toole, however, enabled users to proactively align intent before generation—typically after reviewing the entire intent tree in the first round—thereby reducing the number of iterations. For example, P6 commented, \textit{``I can solve many problems at once before generating code, which is really convenient compared with using ChatGPT directly.''} In later rounds, the simplified graph further accelerated instruction delivery.}

\rv{\textbf{Changed Debugging and Testing Behaviors.} We observed how \tool changed developer behaviors in testing and debugging. First, users could perform code-free, task-directed testing and debugging, directly interacting with subtasks without needing to understand or modify code, as required in traditional program development. Participants (\eg, P7, P11) reported that understanding bugs and testing outcomes became less difficult, allowing them to focus more on the task itself. Second, by leveraging the understanding graph, they shifted from sequential, step-by-step debugging and testing to parallel, one-shot processes. Participants noted that compressing multi-round debugging and testing improved efficiency.}

\textbf{Limitations and Suggestions.}
Nevertheless, a few limitations were also reported. Some participants (\eg P1, P10) found that the initial learning curve of the graph interface was relatively steep, especially for users without prior exposure to task-structured programming. 
In addition, participants held differing opinions on the appropriate level of detail in graph nodes. While some preferred concise and abstract task descriptions, others expressed the need for more technical detail, such as variable names and code-level semantics (P6, P7). Thus, they suggested providing customizable levels of granularity and tutorial support to accommodate their diverse backgrounds and preferences.

\rv{\textbf{Design Takeaways.} Based on the above qualitative insights, we distill two design takeaways for future systems that wish to externalize LLM Understanding. 
First, since externalization allows users to shift from sequential to parallel misalignment resolution, future systems need to thoughtfully design representations (\eg, task graphs) that consolidate sequential steps in domain-specific workflows like programming. Second, direct task-intent matching involves multi-round interactions where user intents are constantly changing. The system should provide targeted information aligned with updated user intents to reduce users’ cognitive load and improve system usability.}

\section{Discussion}
We reflect on the broader implications of our approach, potential for generalization, and directions for future improvement.

\subsection{Towards Personalized LLM Task Representations}

While the graph-based representation in \tool has been shown to support effective task viewing, tracking, and editing, our user study revealed considerable variability in user preferences regarding how such representations should be presented. 
Specifically, participants expressed differing needs around the level of granularity (\eg whether nodes should encapsulate high-level concepts or detailed operations), the inclusion of domain-specific metadata (\eg Python library names), and the spatial layout of task graphs. The current system adopts a uniform design grounded in formative study findings, but does not yet support individual customization. Enabling personalization could further lower the interaction threshold and increase system transparency across diverse skill levels.

To achieve this, adaptive strategies such as active learning offer promising potential \cite{activelearningsurvey}. By continuously collecting feedback from user interactions, such as graph edits, task confirmations, and exploration behaviors, the system could learn users' preferred presentation styles and task framing patterns. Over time, this would allow the graph interface to evolve toward more user-aligned views, enhancing usability and interpretability. Future work could explore fine-grained user modeling and incremental interface adaptation to realize personalized task understanding at scale.

\subsection{Beyond Code: Generalizing the Paradigm}

Although this work focuses on code generation, the core concept of externalizing LLM understanding and aligning it with user intent can be extended to a wide range of complex reasoning tasks. In domains such as writing assistance~\cite{Zhang2023b}, data analysis~\cite{waitgpt}, data visualization~\cite{NLISurvey}, and creative design~\cite{dataplaywright, DVSurvey}, users often face similar challenges of intent drift, semantic ambiguity, and nonlinear task structures. The graph-based intermediate layer proposed in \tool offers a general mechanism for making LLM reasoning more accessible and editable, supporting iterative refinement across diverse contexts.

In addition, for the NLP community, our findings suggest new directions for aligning LLMs with human goals, particularly through intent-structured feedforward representations and task-aware input conditioning. Rather than optimizing whole output sequences, users can adjust high-level logic directly through graph manipulation, potentially enabling more efficient feedback collection and targeted preference learning, such as via RLHF or DPO~\cite{nlp_align}. In HCI, this work contributes to the broader conversation on feedforward mechanisms~\cite{min2025feedforward}, demonstrating how intermediate representations can reduce interaction overhead and cognitive burden during co-creative workflows.

Moreover, the proposed paradigm of \textit{direct intent–task matching} holds promise for real-world deployment. Its lightweight, plug-in-style implementation makes it suitable for integration with major LLM platforms, providing non-technical users with a more structured and controllable way to complete tasks. In educational settings such as programming literacy or STEM learning~\cite{NotePlayer}, this paradigm may also help cultivate procedural thinking by shifting attention from code syntax to task logic, an avenue that merits further exploration.

\subsection{Limitations and Future Work}

\textbf{Limitations.} Despite its advantages, \tool also has several limitations. First, efficiency remains a practical concern. 
The current pipeline introduces latency (10–15 seconds) when generating and updating the understanding graph, which may hinder the fluidity of interaction. 
Improving backend processing efficiency through caching, incremental updates, or lightweight modeling may help reduce user wait time and improve responsiveness.
Second, \tool is effective for single-task alignment, it is less suited to evolving multi-task scenarios where user goals shift or expand over time. Thus, \rv{complex scenarios, such as large-scale multi-file projects, were not covered in our user study and require further investigation.} \rv{Lastly, we evaluated the quality of the triples generated by fine-tuned SLMs based on user feedback, which involved users directly examining the understanding graphs in Panel B and the user intent and mappings through the simplified graph in Panel C. While users provided positive feedback on the triples, future work could include a direct technical assessment of triple quality, requiring benchmarks and graph evaluation methods.
}

\textbf{Future work.} First, \tool currently injects the adjusted LLM understanding as textual descriptions. Future work could investigate more advanced integration mechanisms, such as incorporating task representations into model embeddings or prompts via structured schema or feature vectors, to better preserve semantic intent. Second, \rv{though \tool offers benefits for debugging and testing, it also introduces some challenges when developing software without writing code. For example, NeuroSync mainly focuses on solving task-level bugs while leaving code-level bugs for users to solve manually; combining different levels of debugging requires further exploration. Additionally, ``code without code'' may limit users’ long-term coding skills. NeuroSync could be improved by adding features to help users learn to code and prepare for debugging in large codebases.}

\section{Conclusion}

We address the problem of \textit{bidirectional ambiguity} in conversational LLM programming for non-professional users.
To resolve this, we propose \textit{direct intent–task matching}, a new paradigm that externalizes LLM understanding for direct user inspection and editing before code generation. We realize this approach in \textit{\toole}, a system that combines visual representations, graph simplification, and distillation-based efficient extraction to support alignment. Through technical evaluations and a user study, we show that \tool improves alignment, reduces cognitive load, and enhances coding efficiency, offering a promising direction for more transparent and accessible human–LLM collaboration.

\begin{acks}
The authors would like to thank the reviewers for their constructive feedback. The authors also wish to thank Liwenhan Xie, Bopei Nie, Jason Wong, Rui Sheng and Yanna Lin for their advice and support.  This work was supported by the RGC
GRF Grant 16218724.
\end{acks}

\bibliographystyle{ACM-Reference-Format}
\bibliography{sample-base}



\end{document}